\documentclass[aps,prl,twocolumn, superscriptaddress,showpacs,preprintnumbers,amsmath,amssymb]{revtex4}
\usepackage{color}
\usepackage{longtable}
\usepackage{graphicx} % Include figure files
\usepackage{dcolumn}  % Align table columns on decimal point
\usepackage{multirow}

\graphicspath{{ps}}

\def\Journal#1#2#3#4{{#1} {\bf #2}, #3 (#4)}

\def\PLB{{ Phys. Lett.}  B}
\def\PRL{ Phys. Rev. Lett.}
\def\PRD{{ Phys. Rev.} D}

\def\chicx{\chi_{cJ}}
\def\NIMA{{ Nucl. Instrum. Methods Phys. Res., Sect. A }}
\def\NPB{{Nucl. Phys. B}}
\def\pccm{p^*_{\chi_{c\rm{J}}}}

\begin{document}

%\preprint{\vbox{
%    \hbox{Version 0.34} \\ 
%    \hbox{Belle Preprint 2015-xx} \\
%    \hbox{KEK   Preprint 2015-xx} \\
%    \hbox{hep-ex/15-xxxx} }}
    \title{\quad\\[0.5cm] Inclusive and exclusive measurements of $B$ decays to $\chi_{c1}$ and $\chi_{c2}$ at Belle }

\affiliation{University of the Basque Country UPV/EHU, 48080 Bilbao}
\affiliation{Beihang University, Beijing 100191}
%%%\affiliation{University of Bonn, 53115 Bonn}
\affiliation{Budker Institute of Nuclear Physics SB RAS, Novosibirsk 630090}
\affiliation{Faculty of Mathematics and Physics, Charles University, 121 16 Prague}
%%%\affiliation{Chiba University, Chiba 263-8522}
\affiliation{Chonnam National University, Kwangju 660-701}
\affiliation{University of Cincinnati, Cincinnati, Ohio 45221}
\affiliation{Deutsches Elektronen--Synchrotron, 22607 Hamburg}
%%%\affiliation{University of Florida, Gainesville, Florida 32611}
%%%\affiliation{Department of Physics, Fu Jen Catholic University, Taipei 24205}
\affiliation{Justus-Liebig-Universit\"at Gie\ss{}en, 35392 Gie\ss{}en}
\affiliation{Gifu University, Gifu 501-1193}
\affiliation{II. Physikalisches Institut, Georg-August-Universit\"at G\"ottingen, 37073 G\"ottingen}
\affiliation{SOKENDAI (The Graduate University for Advanced Studies), Hayama 240-0193}
\affiliation{Gyeongsang National University, Chinju 660-701}
\affiliation{Hanyang University, Seoul 133-791}
\affiliation{University of Hawaii, Honolulu, Hawaii 96822}
\affiliation{High Energy Accelerator Research Organization (KEK), Tsukuba 305-0801}
%%%\affiliation{Hiroshima Institute of Technology, Hiroshima 731-5193}
\affiliation{IKERBASQUE, Basque Foundation for Science, 48013 Bilbao}
%%%\affiliation{University of Illinois at Urbana-Champaign, Urbana, Illinois 61801}
\affiliation{Indian Institute of Technology Bhubaneswar, Satya Nagar 751007}
\affiliation{Indian Institute of Technology Guwahati, Assam 781039}
\affiliation{Indian Institute of Technology Madras, Chennai 600036}
%%%\affiliation{Indiana University, Bloomington, Indiana 47408}
\affiliation{Institute of High Energy Physics, Chinese Academy of Sciences, Beijing 100049}
\affiliation{Institute of High Energy Physics, Vienna 1050}
\affiliation{Institute for High Energy Physics, Protvino 142281}
%%%\affiliation{Institute of Mathematical Sciences, Chennai 600113}
\affiliation{INFN - Sezione di Torino, 10125 Torino}
%%%\affiliation{Institute for Theoretical and Experimental Physics, Moscow 117218}
\affiliation{J. Stefan Institute, 1000 Ljubljana}
\affiliation{Kanagawa University, Yokohama 221-8686}
\affiliation{Institut f\"ur Experimentelle Kernphysik, Karlsruher Institut f\"ur Technologie, 76131 Karlsruhe}
%%%\affiliation{Kavli Institute for the Physics and Mathematics of the Universe (WPI), University of Tokyo, Kashiwa 277-8583}
\affiliation{Kennesaw State University, Kennesaw GA 30144}
\affiliation{King Abdulaziz City for Science and Technology, Riyadh 11442}
\affiliation{Department of Physics, Faculty of Science, King Abdulaziz University, Jeddah 21589}
\affiliation{Korea Institute of Science and Technology Information, Daejeon 305-806}
\affiliation{Korea University, Seoul 136-713}
%%%\affiliation{Kyoto University, Kyoto 606-8502}
\affiliation{Kyungpook National University, Daegu 702-701}
\affiliation{\'Ecole Polytechnique F\'ed\'erale de Lausanne (EPFL), Lausanne 1015}
\affiliation{Faculty of Mathematics and Physics, University of Ljubljana, 1000 Ljubljana}
\affiliation{Ludwig Maximilians University, 80539 Munich}
\affiliation{Luther College, Decorah, Iowa 52101}
\affiliation{University of Maribor, 2000 Maribor}
\affiliation{Max-Planck-Institut f\"ur Physik, 80805 M\"unchen}
\affiliation{School of Physics, University of Melbourne, Victoria 3010}
\affiliation{Middle East Technical University, 06531 Ankara}
\affiliation{Moscow Physical Engineering Institute, Moscow 115409}
\affiliation{Moscow Institute of Physics and Technology, Moscow Region 141700}
\affiliation{Graduate School of Science, Nagoya University, Nagoya 464-8602}
\affiliation{Kobayashi-Maskawa Institute, Nagoya University, Nagoya 464-8602}
%%%\affiliation{Nara University of Education, Nara 630-8528}
\affiliation{Nara Women's University, Nara 630-8506}
\affiliation{National Central University, Chung-li 32054}
\affiliation{National United University, Miao Li 36003}
\affiliation{Department of Physics, National Taiwan University, Taipei 10617}
\affiliation{H. Niewodniczanski Institute of Nuclear Physics, Krakow 31-342}
%%%\affiliation{Nippon Dental University, Niigata 951-8580}
\affiliation{Niigata University, Niigata 950-2181}
%%%\affiliation{University of Nova Gorica, 5000 Nova Gorica}
\affiliation{Novosibirsk State University, Novosibirsk 630090}
\affiliation{Osaka City University, Osaka 558-8585}
%%%\affiliation{Osaka University, Osaka 565-0871}
\affiliation{Pacific Northwest National Laboratory, Richland, Washington 99352}
\affiliation{Panjab University, Chandigarh 160014}
\affiliation{Peking University, Beijing 100871}
\affiliation{University of Pittsburgh, Pittsburgh, Pennsylvania 15260}
\affiliation{Punjab Agricultural University, Ludhiana 141004}
%%%\affiliation{Research Center for Electron Photon Science, Tohoku University, Sendai 980-8578}
%%%\affiliation{Research Center for Nuclear Physics, Osaka University, Osaka 567-0047}
%%%\affiliation{RIKEN BNL Research Center, Upton, New York 11973}
%%%\affiliation{Saga University, Saga 840-8502}
\affiliation{University of Science and Technology of China, Hefei 230026}
\affiliation{Seoul National University, Seoul 151-742}
%%%\affiliation{Shinshu University, Nagano 390-8621}
\affiliation{Soongsil University, Seoul 156-743}
\affiliation{University of South Carolina, Columbia, South Carolina 29208}
\affiliation{Sungkyunkwan University, Suwon 440-746}
\affiliation{School of Physics, University of Sydney, NSW 2006}
\affiliation{Department of Physics, Faculty of Science, University of Tabuk, Tabuk 71451}
\affiliation{Tata Institute of Fundamental Research, Mumbai 400005}
\affiliation{Excellence Cluster Universe, Technische Universit\"at M\"unchen, 85748 Garching}
\affiliation{Department of Physics, Technische Universit\"at M\"unchen, 85748 Garching}
\affiliation{Toho University, Funabashi 274-8510}
%%%\affiliation{Tohoku Gakuin University, Tagajo 985-8537}
\affiliation{Tohoku University, Sendai 980-8578}
\affiliation{Earthquake Research Institute, University of Tokyo, Tokyo 113-0032}
\affiliation{Department of Physics, University of Tokyo, Tokyo 113-0033}
\affiliation{Tokyo Institute of Technology, Tokyo 152-8550}
\affiliation{Tokyo Metropolitan University, Tokyo 192-0397}
%%%\affiliation{Tokyo University of Agriculture and Technology, Tokyo 184-8588}
\affiliation{University of Torino, 10124 Torino}
%%%\affiliation{Toyama National College of Maritime Technology, Toyama 933-0293}
\affiliation{Utkal University, Bhubaneswar 751004}
\affiliation{CNP, Virginia Polytechnic Institute and State University, Blacksburg, Virginia 24061}
\affiliation{Wayne State University, Detroit, Michigan 48202}
\affiliation{Yamagata University, Yamagata 990-8560}
\affiliation{Yonsei University, Seoul 120-749}
\author{V.~Bhardwaj}\affiliation{University of South Carolina, Columbia, South Carolina 29208} % SouthCarolina 
\author{K.~Miyabayashi}\affiliation{Nara Women's University, Nara 630-8506} % Nara
\author{E.~Panzenb\"ock}\affiliation{II. Physikalisches Institut, Georg-August-Universit\"at G\"ottingen, 37073 G\"ottingen}\affiliation{Nara Women's University, Nara 630-8506} % Goettingen
\author{K.~Trabelsi}\affiliation{High Energy Accelerator Research Organization (KEK), Tsukuba 305-0801}\affiliation{SOKENDAI (The Graduate University for Advanced Studies), Hayama 240-0193} % KEK
 \author{A.~Frey}\affiliation{II. Physikalisches Institut, Georg-August-Universit\"at G\"ottingen, 37073 G\"ottingen} % Goettingen

 \author{A.~Abdesselam}\affiliation{Department of Physics, Faculty of Science, University of Tabuk, Tabuk 71451} % Tabuk
  \author{I.~Adachi}\affiliation{High Energy Accelerator Research Organization (KEK), Tsukuba 305-0801}\affiliation{SOKENDAI (The Graduate University for Advanced Studies), Hayama 240-0193} % KEK
% \author{K.~Adamczyk}\affiliation{H. Niewodniczanski Institute of Nuclear Physics, Krakow 31-342} % Krakow
  \author{H.~Aihara}\affiliation{Department of Physics, University of Tokyo, Tokyo 113-0033} % Tokyo
  \author{S.~Al~Said}\affiliation{Department of Physics, Faculty of Science, University of Tabuk, Tabuk 71451}\affiliation{Department of Physics, Faculty of Science, King Abdulaziz University, Jeddah 21589} % Tabuk
  \author{K.~Arinstein}\affiliation{Budker Institute of Nuclear Physics SB RAS, Novosibirsk 630090}\affiliation{Novosibirsk State University, Novosibirsk 630090} % BINP
% \author{Y.~Arita}\affiliation{Graduate School of Science, Nagoya University, Nagoya 464-8602} % Nagoya
% ...
% Ljubljana
% \author{N.~Zwahlen}\affiliation{\'Ecole Polytechnique F\'ed\'erale de Lausanne (EPFL), Lausanne 1015} % Lausanne
% \author{O.~Zyukova}\affiliation{Budker Institute of Nuclear Physics SB RAS, Novosibirsk 630090}\affiliation{Novosibirsk State University, Novosibirsk 630090} % BINP
%\collaboration{The Belle Collaboration}
 \collaboration{The Belle Collaboration}

\begin{abstract}
  We report inclusive and exclusive measurements for 
  $\chi_{c1}$ and $\chi_{c2}$ production in $B$ decays. We measure 
  $\mathcal{B}(B \to \chi_{c1} X)$=
  $(3.03 \pm 0.05(\mbox{stat}) \pm 0.24(\mbox{syst})) \times 10^{-3}$  and
  $\mathcal{B}(B \to \chi_{c2} X)$=
  $(0.70 \pm 0.06(\mbox{stat}) \pm 0.10(\mbox{syst})) \times 10^{-3}$.
  For the first time, $\chi_{c2}$ production in 
  exclusive $B$ decays in the modes $B^0 \to \chi_{c2}\pi^- K^+$
  and  $B^+ \to \chi_{c2} \pi^+ \pi^- K^+$ has been observed, along 
  with first evidence for the 
  $B^+ \to \chi_{c2} \pi^+ K_S^0$ decay mode. For $\chi_{c1}$ production,
  we  report the  first observation in the 
  $B^+ \to \chi_{c1} \pi^+ \pi^- K^+$,
  $B^0 \to \chi_{c1}  \pi^+ \pi^- K_S^0$  and 
  $B^0 \to \chi_{c1}  \pi^0 \pi^- K^+$ decay modes. Using these decay modes, we
  observe a difference in the   production mechanism of $\chi_{c2}$ in 
  comparison to $\chi_{c1}$ in $B$ decays. 
  In addition, we report searches for $X(3872)$ and $\chi_{c1}(2P)$ in the
  $B^+ \to (\chi_{c1} \pi^+ \pi^-) K^+$ decay mode.
  The reported results use 
  $772 \times 10^{6}$ $B\overline{B}$ events collected at the 
  $\Upsilon(4S)$ resonance with the Belle detector at the KEKB 
  asymmetric-energy $e^+e^-$ collider.
\end{abstract}


\maketitle

%.................
%    INTRODUCTION
%................

\section{Introduction}
%A decade has passed since 
Belle reported the first observation of 
$\chi_{c2}$ production in $B$ meson decays with an inclusive 
measurement~\cite{Belle_PRL_89_011803_2002}.
The $\chicx (J=1,2)$~\cite{CX} momentum  distributions in the 
$\Upsilon(4S)$ rest frame (CM frame)  indicate that most of the
$\chi_{c2}$ mesons come from  non-two-body $B$ 
decays~\cite{Belle_PRL_89_011803_2002, BaBar_PRD_67_032002_2003}. 
Still, there have been only a few
searches for exclusive $B$ decays with a $\chi_{c2}$ in the 
final state, $B^+ \to \chi_{c2} K^+$~\cite{Belle_PRL_107_091803_2011} and 
$B^0 \to \chi_{c2}  K^*(892)^0$~\cite{Belle_Soni_2006,BaBar_PRL_102_132001_2009,LHCb_NPB_874_3_663_2013}.  
The $B^+ \to \chi_{c2} K^{(*)}$ decays are found to be 
highly suppressed with respect to 
the similar $\chi_{c1}$ processes~\cite{pdg}.
The suppression can be explained in the framework of the
factorization in two-body $B$ decays~\cite{bdecays_fact}, 
where $\chi_{c2}$ production is allowed  only when 
one takes into account final state interactions. 
%which is based on small final state interaction.
%As the $B$ meson mass is relatively large, two decay 
%daughter particles go back to back in the $B$ meson rest frame with high 
%momentum and move quickly apart from each other. 
Due to angular momentum conservation, $J^{PC}=0^{-+}$, 
$1^{--}$ and $1^{++}$ are favored while $0^{++}$, $2^{++}$, $2^{--}$ and so on are
suppressed. 

A study of the multi-body $B$ decay modes with $\chi_{c1}$ and $\chi_{c2}$
in the final state is important to understand 
the detailed dynamics of $B$ meson decays.
Further, one can search for charmonium/charmonium-like exotic states in
one of the intermediate final states such as $\chicx \pi$ and $\chicx \pi \pi$. 
For example, looking at the $\chi_{c1} \pi^+\pi^-$ invariant mass
spectrum in $B \to \chi_{c1} \pi^+\pi^- K$ decays,  one can search for
$\chi_{c1}(2P)$ and/or $X(3872)$. The quantum numbers of the 
narrow exotic resonance $X(3872)$ have been determined to be 
$J^{PC}=1^{++}$~\cite{CDF_PRL_98_132002,Belle_PRD_84_052004,LHCb_PRL_110_222001}.
One plausible interpretation is an
admixture of a $D^0 \bar{D}^{*0}$ molecule and a conventional
charmonium with the same $J^{PC}$, the yet-unseen $\chi_{c1}(2P)$~\cite{ccddmix}.
The $\chi_{c1}(2P)$ component may have a substantial decay rate to
$\chi_{c1} \pi^+ \pi^-$ because of no obvious conflict in
quantum numbers and observations of di-pion transitions between
$\chi_{bJ}$ states in the bottomonium system.
In case  that $X(3872)$ is not a mixed state and hence $\chi_{c1}(2P)$
is a physically observable state, its decay to
$\chi_{c1} \pi^+\pi^-$ would still be expected. Its mass is predicted
to be about 3920 MeV/$c^2$, 
assuming that it lies between $\chi_{c2}(2P)$
and the $X(3915)$ that is interpreted as  $\chi_{c0}(2P)$  by PDG~\cite{pdg}. 

Using the $\chicx \to J/\psi \gamma$ modes, we report on the 
inclusive branching fractions ($\mathcal{B}$) of $B \to \chicx X$ 
decays and the exclusive reconstruction of multi-body $B$ decays 
to $\chicx$ in order to search for still-undiscovered
intermediate states.

%...................
%  Data sample and detector
%

\section{Data sample and detector}
We use a data sample of $772 \times 10^{6}$ $B\bar{B}$ events collected
%$711$ fb$^{-1}$ collected 
with the Belle detector~\cite{abashian} at the KEKB asymmetric-energy 
$e^+e^-$ collider operating at the $\Upsilon(4S)$ resonance~\cite{kurokawa}. 
The Belle detector is a large-solid-angle spectrometer, which includes a
silicon vertex detector (SVD), a 50-layer central drift chamber (CDC),
an array of aerogel threshold Cherenkov counters (ACC), time-of-flight
scintillation counters (TOF), and an electromagnetic calorimeter (ECL) 
comprised of 8736 CsI(Tl) crystals located inside a superconducting 
solenoid coil that provides a 1.5~T magnetic field. An iron flux return 
located outside the coil is instrumented to detect $K^{0}_{L}$ mesons and 
identify muons (KLM). 
The detector is described in detail elsewhere~\cite{abashian}. 
 Two inner detector configurations were used. 
A first sample of $152 \times 10^{6}$ $B\bar{B}$ events was collected with 
a 2.0~cm radius beam-pipe and a 3-layer SVD, while the remaining
$620 \times 10^{6}$ $B\bar{B}$ pairs were collected with a 1.5 cm 
radius beam pipe, a 4-layer silicon detector and  modified CDC 
  (the cathode part of the CDC replaced by a compact  small 
cell-type drift chamber)~\cite{SVD2_NIMA_560_1_2006}.

%%%%%%
\section{Event selection}

We  reconstruct inclusive $\chicx$ from $B$ decays.  To
suppress continuum background, we exploit the $\Upsilon(4S)$ 
decay topology. For the events passing the Belle standard hadronic event 
selection~\cite{belle_b2cc2002}, we require the ratio of the second to
zeroth Fox-Wolfram moment~\cite{FoxWolf} to be less than 0.5.
Charged tracks are required to originate from  
the vicinity of the interaction point (IP): the distance of closest approach 
to the IP
is  required to be within 3.5~cm  along the beam
direction and within 
1.0~cm in the transverse plane. Photons are reconstructed 
from  the energy deposition in the ECL by requiring no matching with any 
extrapolated charged track.   To further avoid photons coming 
from neutral hadrons, we reject the photon candidate if the 
ratio of the energy deposited in the central array of 3$\times$3 ECL cells 
to that deposited in the enclosing array of 5$\times$5 cells is less than 0.85.

We use EVTGEN~\cite{EvtGen} with QED final state radiation by 
PHOTOS~\cite{PHOTOS} for the generation of
Monte-Carlo (MC) simulation events.   
A GEANT-based~\cite{GEANT} MC simulation is used to model the response
of the detector 
and determine the efficiency of the signal reconstruction.

The $J/\psi$ meson is reconstructed via its decays to 
$\ell^+ \ell^-$ ($\ell$ = $e$ or $\mu$) and selected by the invariant
mass $M_{\ell\ell}$. 
 For the di-muon mode,  $M_{\ell \ell}$ is given by the
invariant mass  $M_{\mu^+\mu^-}$; for 
the di-electron mode,
the four-momenta of all photons within 50 mrad with respect to the
original direction of the $e^+$ or $e^-$ tracks are included in 
 $M_{\ell \ell} \equiv M_{e^+e^-(\gamma)}$ to reduce the radiative tail.
The reconstructed invariant mass of the $J/\psi$ candidates is required 
to satisfy 2.95~GeV$/c^2 < M_{e^+ e^-(\gamma)} < 3.13$~GeV$/c^2$ or 
3.03~GeV$/c^2 < M_{\mu^+ \mu^-} < 3.13$~GeV$/c^2$. 
For the selected $J/\psi$ candidates, a vertex-constrained fit is
applied to the charged tracks and then a mass-constrained fit
is performed to improve its momentum resolution.  The $\chi_{c1}$ and $\chi_{c2}$ candidates are 
reconstructed by combining a $J/\psi$ candidate with a photon having an energy
larger than 100~MeV.

\section{Inclusive $B$ decays to $\chicx$}
%
%\subsection{$\chicx$ reconstruction}

\subsection{Branching fraction measurement}

To reduce  combinatorial
background coming from $\pi^0 \to \gamma\gamma$, we use a likelihood function
that distinguishes an isolated photon from $\pi^0$ decays using the photon-pair
invariant mass, the photon laboratory-frame energy, and the laboratory-frame polar angle 
with respect to the beam direction~\cite{koppenburg}.  
We reject both photons of a pair whose $\pi^0$ likelihood probability 
is larger than 0.3.  Applying this cut, combinatorial background is reduced by 
56.9\% (59.1\%) with a signal loss of 26.5\% (39.9\%) for $\chi_{c2}$ 
($\chi_{c1}$).

To identify the signal,  we use the distribution of the
$J/\psi \gamma$ invariant mass $M_{J/\psi\gamma}$  and
extract the signal yield from a binned  
maximum likelihood fit. %the variable $M_{J/\psi \gamma}$.
The signal of $\chicx$ is described by a double-sided Crystal Ball 
function~\cite{crystal_ball,dCB}, which accommodates
the tails of the mass distribution. The function's left (right)
side tail parameters $n_l$ ($n_r$) and $\alpha_l$ ($\alpha_r$) 
are fixed to the values obtained from MC simulated events. 
For $B \to \chi_{c1} X$, all other shape parameters are
floated in the fit whereas, for $B \to \chi_{c2} X$, they are fixed using the 
mass difference  ($m_{\chi_{c2}} - m_{\chi_{c1}}$) from Ref.~\cite{pdg} and the 
resolution ratio between  $\chi_{c1}$ and $\chi_{c2}$,
$\sigma_{\chi_{c2}}/\sigma_{\chi_{c1}}$, determined from MC simulations.
The combinatorial background component is modeled with a third-order
Chebyshev polynomial.

%%%

\begin{figure}[h!]
  \centering
  \includegraphics[height=60mm,width=90mm]{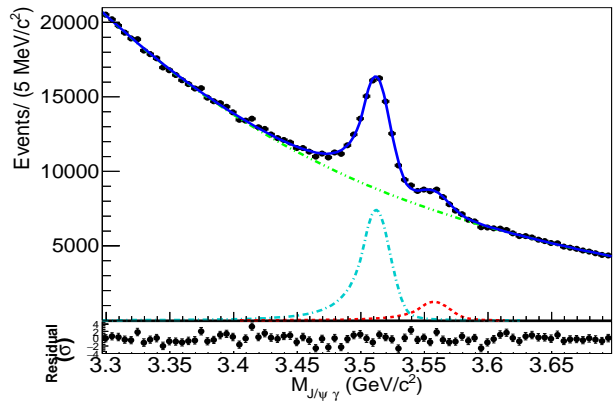}%
  \caption{ (color online) $M_{J/\psi \gamma}$ distribution of the 
       $B \to \chicx(\to J/\psi(\to \ell^+  \ell^-)\gamma) X$ decays in 
       data. The curves show the signal 
       (cyan dash-dotted for $\chi_{c1}$ and red dashed for $\chi_{c2}$)
       and the background component (green dash-double-dotted for 
       combinatorial) as well as the overall fit (blue solid). The lower
       plot shows the pull of the residuals with respect to the fit.}
   \label{fig:incdata_fit}
\end{figure}

Figure~\ref{fig:incdata_fit} shows the fit of the 
$M_{J/\psi \gamma}$ distribution for 
$\chi_{c1} X$ and $\chi_{c2} X$ decays in the range of 
[3.297, 3.697]~GeV$/c^2$.
The fit returns a reduced $\chi^2$ of 1.3 
with a p-value of 0.0123 and
a yield of $51353 \pm 614$ events for the $\chi_{c1}$ 
and $9651 \pm 446$ events for the $\chi_{c2}$, where the errors 
are statistical.

The reconstruction  efficiencies for the inclusive 
$B \to \chi_{c1} X$ and
$B \to \chi_{c2} X$ decays  are estimated to be 24.2\% and 25.9\%, respectively.
The efficiency is estimated using  simulated multi-body $B$ decays,
$B \to \chicx  K (n\pi)$, where the number of pions $n$ 
varies from 0 to 4 over the entire $\pccm$ range; it is averaged with proper weighting
according to the distribution 
of $\pccm$ in data.

We use the 2014 world-average values~\cite{pdg} for secondary
daughter branching fractions
${\mathcal B}(J/\psi\rightarrow l^+l^-) = (11.932\pm 0.004)\%$,
${\mathcal B}(\chi_{c1}\rightarrow J/\psi\gamma) = (33.9\pm 1.2)\%$, and
${\mathcal B}(\chi_{c2}\rightarrow J/\psi\gamma) = (19.2\pm 0.7)\%$.

%Using them, the inclusive $B\rightarrow\chicx X$ branching fractions 
%(statistical uncertainty only) 
%are found to be:
%${\mathcal B}$$(B\rightarrow \chi_{c1}X) = (3.40\pm 0.04)\times 10^{-3}$, and
%${\mathcal B}$$(B\rightarrow \chi_{c2}X) = (1.06\pm 0.05)\times 10^{-3}$,
%respectively. 

We use the 89~$\mbox{fb}^{-1}$ off-resonance data sample taken at 60 MeV below 
the $\Upsilon(4S)$ resonance to estimate the contribution of $\chicx$ 
particles that do not arise from $B$ meson decays. 
From the fit to the $M_{J/\psi \gamma}$ distribution for that sample, 
we obtain $139\pm 38$ ($92\pm 38$) signal events for $\chi_{c1}$ 
($\chi_{c2}$),
%with statistical significance of 3.8$\sigma$ (2.4$\sigma$).  
corresponding to  $1098\pm 300$ ($727\pm 300$)  signal events for
$\chi_{c1} X$ ($\chi_{c2} X$)  after proper scaling to the integrated 
luminosity at the $\Upsilon(4S)$ resonance.
The scaled $\chi_{c1}$ and $\chi_{c2}$ continuum yields are 
subtracted from the on-resonance yields.

One also expects a contribution from ``feed-down''
$B\to \chicx X$ decays 
where the $\chicx$ is from the cascade 
$B\to \psi' X \to \chicx \gamma X$.
%, we call this contribution ``feed-down'' throughout this paper.
 To determine the rate for direct decays to 
the $\chicx$ states, we subtract this feed-down
contribution, which is estimated using 
$\mathcal{B}(B\rightarrow \psi' X)$ and 
$\mathcal{B}(\psi' \rightarrow \chicx \gamma)$ from Ref.~\cite{pdg}.
%The branching fractions are summarized in Table~\ref{tab:yields}.

%\begin{table}[htbp!]
%\caption{\label{tab:yields}Branching fractions after subtracting 
%$\psi'$ feed-down and off-resonance continuum yield. First (second) error 
%is statistical (systematic).}
%\begin{tabular}{lcc}
%  \hline \hline  
%Mode & $\mathcal{B}(B \to \chicx X)$ , $10^{-3}$   \\ \hline
%$B \to \chi_{c1} X$  & $3.03\pm 0.04 \pm 0.22$  \\
%$B \to \chi_{c2} X$  & $0.70\pm 0.05 \pm 0.07$ \\
%\hline \hline
%\end{tabular}
%\end{table}

The sources and estimates of the systematic uncertainties 
are summarized in Table~\ref{tab:syserr}.
A correction for small differences in the signal detection efficiency 
between MC and data has been applied for the lepton identification requirements.
Uncertainties  in these corrections are included in the systematic error.  
The  $e^+ e^- \to e^+ e^- \ell^+ \ell^-$   and $J/\psi \to \ell^+ \ell^-$ 
($\ell =$ $e$ or $\mu$) samples are used to estimate the lepton identification 
correction. %~\ref{LIDDD}.
The uncertainty of the probability density function (PDF) shapes are obtained by varying all fixed parameters by
$\pm 1 \sigma$, fitting with different binning, and using a fourth-order 
polynomial for the background, then adding the 
changes in the yield in quadrature 
to get the systematic uncertainty. 
We perform a fit to the data by including the $\chi_{c0}$ component and find 
its statistical significance to be 1.7$\sigma$. We further add the signal yield 
difference for $\chi_{c1}$ or $\chi_{c2}$ with respect to the original fit to the PDF systematic uncertainty.
Based on this, we get an uncertainty 
of 3.1\% (7.9\%) for $B \to \chi_{c1} X$ ($B \to \chi_{c2} X$). 
The uncertainties due to 
the secondary branching fractions are also taken into account.  
The uncertainty on the track finding efficiency is found to be 0.35$\%$ per
track by comparing the data and MC for $D^* \to D^0 \pi$ decay, where 
$D^0 \to \pi^+ \pi^- K_S^0$ and $K_S^0 \to \pi^+ \pi^-$ here one of
the $\pi$ is allowed not to be reconstructed explicitly. 
For $N_{B\bar{B}}$, systematic uncertainty is estimated to be $1.4\%$.
The uncertainty on the photon identification is estimated 
to be 2.0\% from sample of radiative Bhabha events. 
The systematic uncertainty associated with the
difference  of the $\pi^0$ veto  between data and MC 
is estimated to be 1.2\% from  a study of 
the $B^\pm \to \chi_{c1} (\to J/\psi \gamma) K^{\pm}$ sample.
The potential bias to extract signal yields of the $\chicx$ is estimated 
by the MC from variation of the efficiency for the different 
decay modes  bin by bin in the $p^*_{\chicx}$ 
distribution.  The 
efficiency change due to the unknown $\chicx$ polarization is estimated 
using the $B^0 \to \chi_{c1} K^{*0}$ signal MC samples
by varying the polarization 
over the allowed range. The sum of these two effects is  4.0\%.

We measure the feed-down-contaminated branching fractions $\mathcal{B}(B\to \chi_{c1} X)$ and
$\mathcal{B}(B\to \chi_{c2}X)$ to be
$(3.33\pm 0.05\pm 0.24)\times 10^{-3}$ and 
$(0.98\pm0.06\pm0.10)\times 10^{-3}$, respectively, 
where the first (second)
 error is statistical (systematic).
After subtracting the feed-down contribution, we obtain the 
pure inclusive branching fractions 
 $\mathcal{B}(B\to \chi_{c1} X)$ = $(3.03 \pm 0.05\pm 0.24)\times 10^{-3}$ and
 $\mathcal{B}(B\to \chi_{c2}X)$ = $(0.70\pm0.06\pm0.10)\times 10^{-3}$.
In both cases, the systematic uncertainty dominates.
We estimate the inclusive branching fractions according to the formula:
\begin{widetext}
\begin{equation*}
\mathcal{B}(B\to \chicx X) = \frac{N_{\rm sig} - N_{\rm off}}{\epsilon \times N_{B} \times  \mathcal{B}(\chicx \to J/\psi \gamma) \times \mathcal{B}(J/\psi \to \ell^+ \ell^-)} - \mathcal{B}(B \to \psi' X) \times \mathcal{B}(\psi' \to \chicx \gamma)
\end{equation*}
\end{widetext}

Here, $N_{\rm sig}$ is the obtained signal yield, $N_{\rm off}$ is the estimated off-resonance contribution, $\epsilon$ is the reconstruction efficiency,  $N_{B}$ is the number of 
$B$ mesons in the data sample and $\mathcal{B}$ is the branching fraction for
the particular mode taken from~\cite{pdg}. 

\begin{table}[htbp]
\caption{\label{tab:syserr}Summary of systematic uncertainties 
in the $B \to \chicx X$ branching fraction.}

\begin{tabular}{lcc}
 \hline \hline
& \multicolumn{2}{c}{Uncertainty (\%) } \\
Source     & $~~~B \to \chi_{c1} X~~~$  &$~~~B \to \chi_{c2} X~~~$ \\
\colrule
Lepton identification &  2.3         & 2.3      \\
PDF uncertainty & 3.1 & 7.9 \\ 
Secondary ${\mathcal B}$    &  3.6         & 3.7         \\
Tracking efficiency   &  0.7        &  0.7         \\
$N_{B\bar{B}}$ & 1.4 & 1.4 \\
Photon efficiency     &  2.0         & 2.0         \\
$\pi^0$ veto  & 1.2 & 1.2 \\
$B\to\chicx X$ modeling & 4.0 & 4.0 \\
$\psi'$ feed-down &    1.0 & 3.0 \\
\colrule
Total                 &  7.3      & 10.7       \\
\hline \hline
\end{tabular}

\end{table}

%%%
\begin{figure}[h!]
  \centering
  \includegraphics[trim=0.5cm 0.6cm 0.0cm 0.0cm,height=60mm,width=90mm]{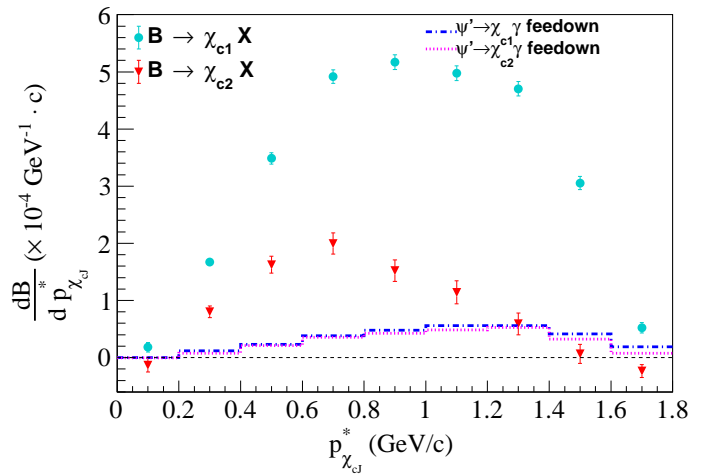}
%{./off_data_bin_all.eps}
  \caption{{(color online) Differential branching fractions
      $D\mathcal{B}(B \to \chi_{c1} X)$ with cyan circles 
      ({\color{cyan} $\bullet$}) and 
       $D\mathcal{B}(B\to\chi_{c2}X)$ with red 
      triangles   ({\color{red} $\blacktriangledown$}) in each bin of $\pccm$ 
      extracted from the maximum likelihood (ML)
      fits performed with 
      the $\Delta M$ 
      distribution of the  data sample for 
      $B \to \chicx(\to J/\psi\gamma) X$ after continuum  suppression 
      but without feed-down subtraction. 
      The $\psi'$ feed-down component (estimated from the MC simulation 
      using $\psi'$ momentum distribution presented in 
      Ref.~\cite{BaBar_PRD_67_032002_2003}) is shown by blue 
      dash-dotted (magenta dotted) 
      line for $\chi_{c1}$ ($\chi_{c2}$).
      The uncertainties in 
      these plots are statistical only.}}
  \label{fig:datainc_fit_both}
\end{figure}

The ratio 
$\mathcal{R_{B}}\equiv \mathcal{B}(B\to \chi_{c2} X)/\mathcal{B}(B\to \chi_{c1}X)$ is $(23.1 \pm 2.0 \pm 2.1)\%$. Here, most of the 
systematics cancel  except for the PDF uncertainty ($4.5\%$), 
secondary $\mathcal{B}$ ($4.7\%$),
unknown polarization ($5.6\%$), and  feed-down ($2.1\%$).

\subsection{ {\boldmath$\pccm$} distribution}
The distribution of the $\chicx$ momentum in the $e^+e^-$ 
center-of-mass frame, $\pccm$, provides valuable insight into
the production mechanism of  the $\chicx$. 
To obtain the $\pccm$ distribution, we fit the
$M_{J/\psi \gamma}$ distribution in bins of $\pccm$.  We fix all of 
the signal parameters 
to the values obtained from the fit to the total and the resolution in each
bin to the value obtained from the
signal MC after MC/data correction. The background
shape and normalization are floated in all fits. The fitted 
$\chi_{c1}$ and $\chi_{c2}$ yields are converted into differential branching
fractions ($D\mathcal{B}$) after  subtraction of
the continuum contribution in each bin, estimated
from the continuum data. In the absence of 
reliable bin by bin estimation of the feed-down 
contribution, we do not apply feed-down subtraction here.
Efficiency corrections are applied to each bin.
Figure~\ref{fig:datainc_fit_both} shows the resulting distributions
of $D\mathcal{B}$ in bins of $\pccm$.
Suppression of the two-body decay
of $\chi_{c2}$ is visible in the $\pccm$ distribution. Most of the
$\chi_{c2}$  production comes from three- or higher-body decays.

\section{Exclusive reconstruction}
To further understand $\chi_{c1}$ and $\chi_{c2}$ production 
in $B$ decays, 
we reconstruct the following exclusive $B$ decays: 
$B^0 \to \chicx \pi^- K^+$, 
$B^+ \to \chicx \pi^+ K_S^0$, 
$B^+ \to \chicx \pi^0 K^+$, 
$B^+ \to \chicx \pi^+ \pi^- K^+$,
$B^0 \to \chicx \pi^+ \pi^- K_S^0$ and
$B^0 \to \chicx \pi^- \pi^0 K^+$~\cite{mixchg}. 
%$B^0 \to \chi_{c1} \pi^- K^+$, 
%$B^0 \to \chi_{c2} \pi^- K^+$, 
%$B^+ \to \chi_{c1} \pi^+ K_S^0$, 
%$B^+ \to \chi_{c2} \pi^+ K_S^0$, 
%$B^+ \to \chi_{c1} \pi^0 K^+$, 
%$B^+ \to \chi_{c2} \pi^0 K^+$, 
%$B^+ \to \chi_{c1} \pi^+ \pi^- K^+$, 
%$B^+ \to \chi_{c2} \pi^+ \pi^- K^+$, 
%$B^0 \to \chi_{c1} \pi^+ \pi^- K_S^0$ and 
%$B^0 \to \chi_{c2} \pi^+ \pi^- K_S^0$ .

The $\chi_{c1}$ and $\chi_{c2}$ candidates are reconstructed as in 
the inclusive study except for a looser criterion 
to reduce the
$\pi^0 \to \gamma\gamma$ background, requiring the $\pi^0$ likelihood 
probability to be less than 0.8. 
Applying this cut, the combinatorial background is reduced
by 30-35\% with a signal loss of 
6-11\% depending upon the  mode of interest.
The reconstructed invariant 
mass of the $\chi_{c1}$  ($\chi_{c2}$) is required to satisfy 
3.467~GeV$/c^2 < M_{J/\psi\gamma} <$ 3.535~GeV$/c^2$ 
(3.535~GeV$/c^2 < M_{J/\psi \gamma} <$ 3.579~GeV$/c^2$). The selected mass windows
correspond to $[-4.5 \sigma, +2.8\sigma]$ for $\chi_{c1}$ and $[-1.5\sigma, +3.0\sigma]$ for
$\chi_{c2}$ around their nominal mass.
A mass-constrained fit is applied to the selected $\chi_{c1}$ and $\chi_{c2}$ 
candidates.

The combined information from the CDC, TOF and ACC is used to
identify charged kaons and pions based on the $K$/$\pi$ likelihood ratio,
$R_{K}= \mathcal{L}_{K}/(\mathcal{L}_K + \mathcal{L}_\pi)$, where $\mathcal{L}_K$
and $\mathcal{L}_{\pi}$ are likelihood values for the kaon and pion hypotheses,
respectively. A track is identified as a kaon if $R_K$ is greater than
0.6; otherwise, it is classified as
a pion. The kaon (pion) identification 
efficiency lies in the range of $87-94\%$ ($94-97 \%$) while the probability of 
misidentifying a pion (kaon) as a kaon (pion)  is $6.8-10.4\%$ 
($6.5-7.0\%$), depending on the momentum range of kaons and pions. 
To ensure that tracks with low transverse momentum ($p_T$) 
with respect to the beam axis are included
only once as they can curl up and result in duplicate tracks,
criteria similar to those of Refs.~\cite{kakuno,Guler} are 
used:
duplicated tracks for charged pions with $p_T < 0.25$ GeV/$c$ often appear as 
the track pair having $\cos \theta_{\rm open} > 0.95$ 
($\cos \theta_{\rm open} < -0.95$) for same (opposite) charged tracks,
where $\theta_{\rm open}$ is the angle between  the two tracks. 
Among those, when the difference between the absolute value of the 
momentum of the two tracks
is less than 0.1 GeV/$c$, it is treated as a 
duplicate pair. Of the two such
tracks, the one having the closest approach to the
IP is retained.%kept  while the other track is discarded.

$K^{0}_{S}$ mesons are reconstructed by combining two oppositely charged 
pions with an invariant mass $M_{\pi^+\pi^-}$ lying between 
482 and 514~MeV$/c^2$ 
($\pm6\sigma$ around the nominal mass of the $K_S^0$). The selected candidates are 
required to satisfy the quality criteria described in Ref.~\cite{goodks}.
Pairs of photons are combined to form $\pi^0$ candidates within the mass 
range 120~MeV$/c^2 < M_{\gamma\gamma} <$ 150~MeV$/c^2$ ($\pm3\sigma$ around the 
nominal mass of $\pi^0$). 
To reduce combinatorial background, the $\pi^0 \to \gamma \gamma$ candidates 
are also required to have an energy balance parameter 
$|E_1 - E_2|/(E_1 + E_2)$ smaller than 0.8, where $E_1$ ($E_2$) is the energy
of the first (second) photon in the laboratory frame. For each
selected $\pi^0$ candidate, a mass-constrained fit is performed 
to improve its momentum resolution.

To identify the $B$ meson, two kinematic variables are used:
the beam-constrained mass $M_{\rm bc}$ and 
the energy difference  $\Delta E$.
The former is defined as 
$\sqrt{E_{\rm beam}^2/c^2 - (\sum_i \vec{p}_{i})^2}/c$ and the 
latter as $\sum_i E_i - E_{\rm beam}$, where $E_{\rm beam}$ is the beam 
energy in the CM frame and $p_{i}$ ($E_i$) is the momentum (energy) of 
the $i$-th daughter particle in the CM frame; the
summation is over all final-state particles used for
reconstruction. We reject candidates having $M_{\rm bc}$ 
less than 5.27~GeV$/c^2$ or $|\Delta E|~>$ 120~MeV. 
In case of multiple $B$ candidates, we use a statistic $\chi^2$, defined as:
\begin{equation*}
\chi^2 = \chi^2_V + \chi^2_N  + (\frac{ M_{\chicx} - m_{\chicx}}{\sigma_{\chicx}})^2 + 
(\frac{ M_{\rm bc} - m_B}{\sigma_{M_{\rm bc}}})^2, 
\end{equation*}
where $\chi^2_V$  is the reduced $\chi^2$ returned by the vertex fit 
of all
charged tracks, $\chi^2_N$ is  the reduced $\chi^2$ for the $K_S^0$ 
or $\pi^0$ 
mass-constrained fit, $M_{\chicx}$ is the reconstructed mass of $\chicx$, 
and
$m_{\chicx}$ and $m_B$ are the nominal masses of the $\chicx$ and $B$ mesons,
respectively. 
The resolution $\sigma_{M_{\rm bc}}$ of $M_{\rm bc}$, estimated from 
the fit to data, is 3 MeV$/c^2$. The resolution  
$\sigma_{\chi_{c1}}$ ($\sigma_{\chi_{c2}}$) of $\chi_{c1}$  
($\chi_{c2}$), is taken to be 9.5~MeV 
(10.5~MeV) from the inclusive measurements.  The $B$ candidate with 
the lowest $\chi^2$ value is retained. The procedure to select the most
probable $B$ candidate is called best candidate selection (BCS).
After the reconstruction, mean  of 1.1-2.7 $B$ 
candidates per event is found, depending on the decay mode,
 and the BCS chooses the true candidate 
75-98\% of the time.

%\begin{multicols}{2}
\begin{figure*}%[ht]
  %\centering
  \includegraphics[trim=0cm 1cm 0cm 0.25cm,height=90mm,width=160mm]{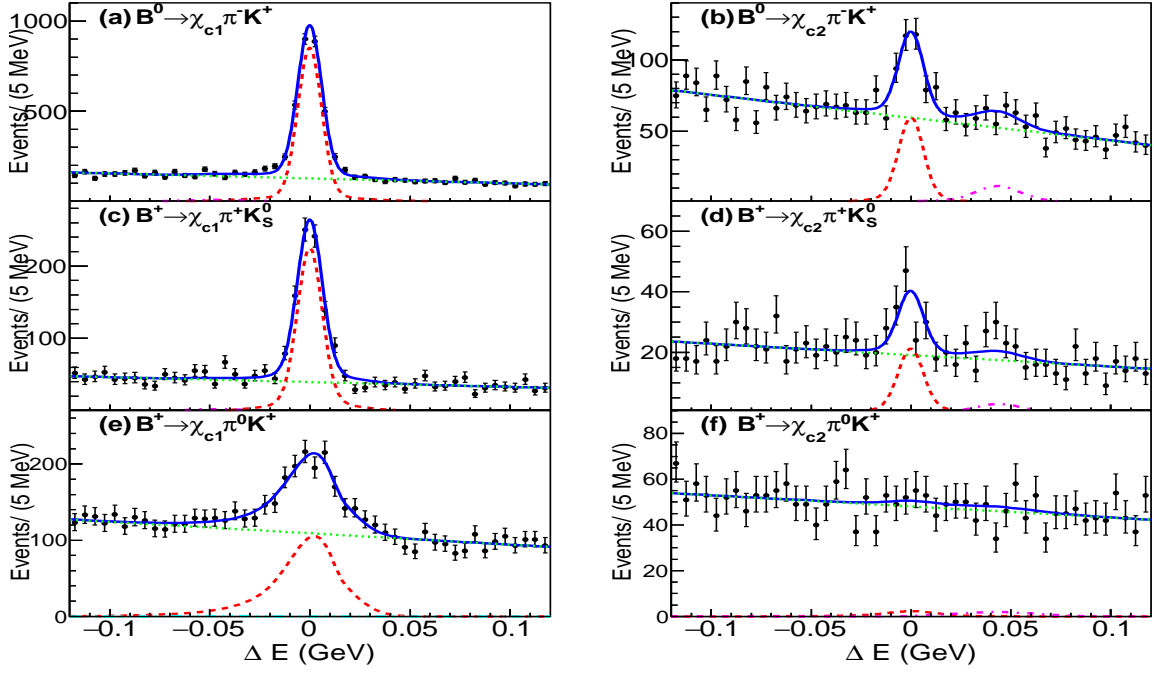}
  \caption{(color online) $\Delta E$ distribution 
    for the (a) $B^0 \to \chi_{c1} \pi^- K^+$,
    (b) $B^0 \to \chi_{c2} \pi^- K^+$,
    (c) $B^+ \to \chi_{c1} \pi^+ K_S^0$,
    (d) $B^+ \to \chi_{c2} \pi^+ K_S^0$,
    (e) $B^+ \to \chi_{c1} \pi^0 K^+$, and
    (f) $B^+ \to \chi_{c2} \pi^0 K^+$ decay modes.
    The curves show the signal 
    (red dashed), the
    peaking background (magenta dash-dotted) 
    and the background component (green dotted for 
    combinatorial) as well as the overall fit (blue solid).}
    \label{fig:B_chicJ_Kpi}
\end{figure*}
%\end{multicols}

 We extract the signal yield from an unbinned 
extended maximum likelihood (UML) fit to the $\Delta E$ variable.
The signal PDF is modeled by the sum of two Gaussians
unless otherwise explicitly 
mentioned. The parameters of the wider Gaussian are fixed
from MC simulations while the mean and the width of the core Gaussian 
are treated according to the $B$ decay mode. 
%floated in the fit (except for the cases limited by statistics {\color{red}
%  which are explicitly mentioned in below)}. 
For the $B\to \chi_{c1} X$ decay modes, the parameters of the 
core Gaussian are floated unless otherwise stated. 
For $B \to \chi_{c2} X$, the core Gaussian is fixed after a data/MC 
correction estimated from  the $B\to \chi_{c1} X$ decay mode;
otherwise, a correction 
from the other decay mode is implemented.

To study the background  from events with a $J/\psi$, we use a large 
MC-simulated
$B \to J/\psi X$ sample corresponding to 100 times 
the integrated luminosity of the data sample. The non-$J/\psi$ (non-$\chicx$) 
background is studied using $M_{\ell\ell}$ ($M_{J/\psi\gamma}$) sidebands 
in data. For $B \to \chi_{c1} X$, no significant peaking background is
found. 
However, in the $B \to \chi_{c2} X$ modes, there can be a 
contamination from $B \to \chi_{c1} X$ because of its larger branching 
fraction.% as well as the limited $J/\psi \gamma$ invariant mass resolution.
We call this effect $B \to \chi_{c1} X$ cross-feed.
Since we apply a mass-constrained
fit for $\chi_{c2} \to  J/\psi \gamma$ candidates, this %the $B \to \chi_{c1}X$ 
cross-feed tends to cluster around $\Delta E = +50$ MeV.
This peaking background is parameterized by a Gaussian whose yield and
parameters are fixed from the signal MC study after applying a MC/data
correction estimated from the $B \to \chi_{c1} X$  decay mode. 
The flat background in all decay modes is 
modeled with a Chebyshev first-order polynomial unless 
otherwise explicitly mentioned. 
For the $B \to \chi_{c1} X$ decay modes, the PDF comprises the signal PDF 
and a flat background; for $B \to \chi_{c2} X$ decay modes, 
the PDF comprises the signal PDF, 
the $B\to \chi_{c1} X$ cross-feed 
and a flat background.

To understand the  intermediate states,
we examine the background-subtracted $M_{\chicx\pi}$,  $M_{K \pi}$, 
$M_{\chicx\pi\pi}$, $M_{K\pi\pi}$, and $M_{\pi\pi}$ distributions for the decay mode of interest.
We perform a UML fit to the $\Delta E$ distribution
and use the $_{S}\mathcal{P}$\textit{lot} 
formalism~\cite{pivk} to project signal events in the distribution.

The efficiency ($\epsilon$) for each decay mode is estimated using 
MC simulation  generated over the whole phase space.
In  the absence of information regarding the intermediate state
and a proper model for each decay mode, we divide the sample according
to the $M_{K n \pi}$ and $M_{\chicx n \pi}$ distributions, where
$n \in \{1,2\}$  is the
number of pions, so that each bin indexed by $i$ 
has equal statistics.
The efficiency estimated in each bin ($\epsilon_i$) using MC simulation is then 
weighted by the signal yield  of the bin to 
provide the final efficiency
$\epsilon = \sum_i w_i \epsilon_i$, where $w_i =$ yield in $i$-th bin /
total yield. In decay modes having no significant signal, the efficiency is
simply estimated using MC simulation  generated over the whole
phase space as distribution is unknown. 
We calibrate this efficiency by the difference
between MC simulation and data, as described
later.  The so-estimated efficiency for the decay mode of interest 
lies between $4.3\%$  and  $18.0\%$, depending upon the final states used for
the reconstruction.
\begin{figure*}%[h!]
  \centering
  \includegraphics[trim=1cm 0.5cm 0.25cm 0.25cm,height=140mm,width=85mm]{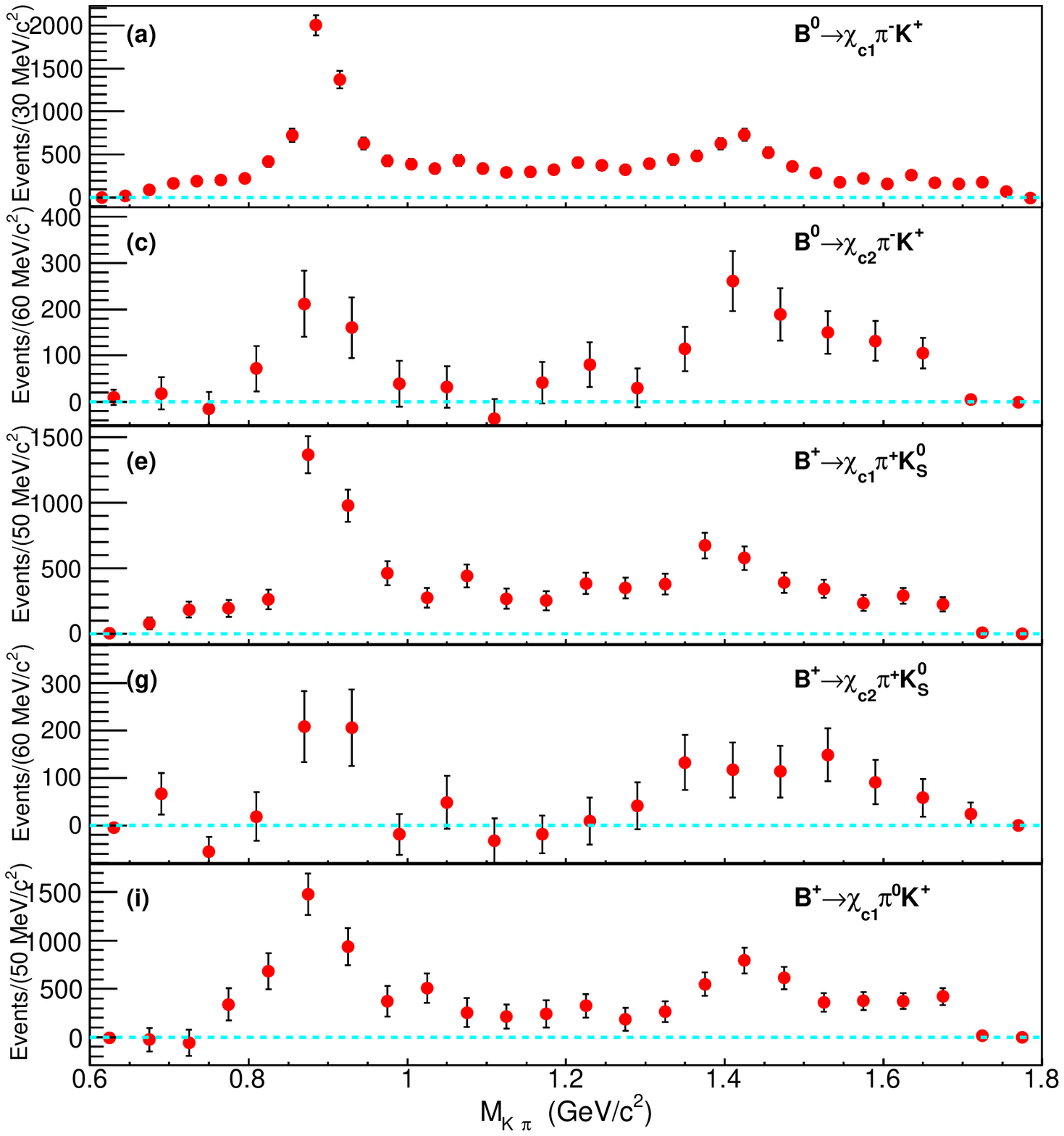} 
  \includegraphics[trim=1.25cm 0.5cm 0.25cm 0.25cm,height=140mm,width=85mm]{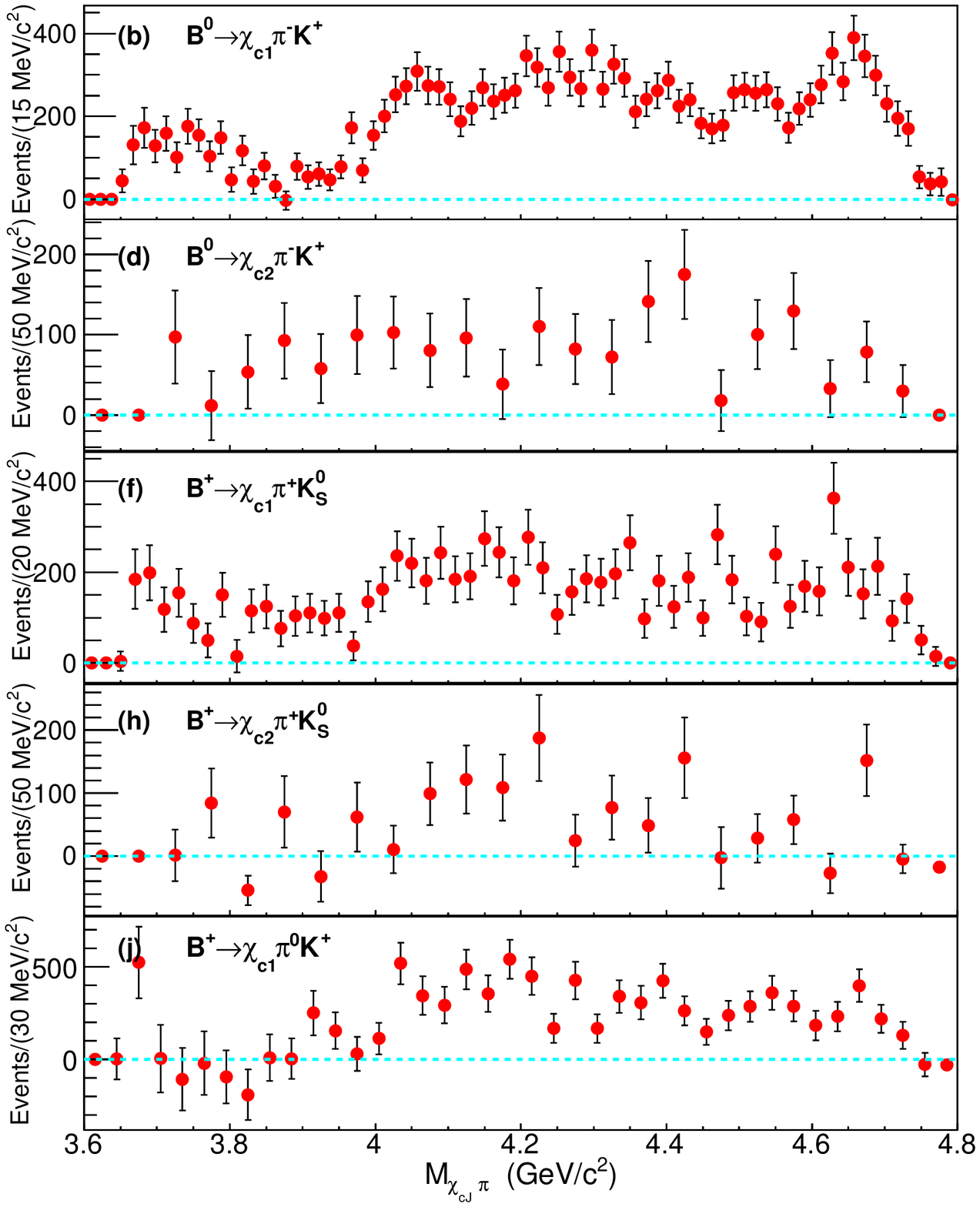}
  
  \caption{(color online) Background subtracted 
    efficiency corrected
    $_{S}\mathcal{P}$\textit{lot}
    $M_{K \pi}$ and $M_{\chicx \pi}$ distributions for the
    (a and b)  $B^0 \to \chi_{c1} \pi^- K^+$,
    (c and d)  $B^0 \to \chi_{c2} \pi^- K^+$,
    (e and f)  $B^+ \to \chi_{c1} \pi^+ K_S^0$,
    (g and h)  $B^+ \to \chi_{c2} \pi^+ K_S^0$ and
    (i and j)  $B^+ \to \chi_{c1} \pi^0 K^+$ decay modes.
  }
  \label{fig:Mchipi_all_3body}
\end{figure*}

%\begin{figure}[h!]
%  \centering
%  \includegraphics[trim=1cm 0.5cm 0cm 0.25cm,height=140mm,width=90mm]{./fig/Pad_mass_Kpi_3body.eps}%Kpimass_threebody.eps}
%  \caption{ Background subtracted $M_{K\pi}$ distribution for
%    (a)  $B^0 \to \chi_{c1} \pi^- K^+$,
%    (b)  $B^0 \to \chi_{c2} \pi^- K^+$,
%    (c)  $B^- \to \chi_{c1} \pi^- K_S^0$,
%    (d)  $B^- \to \chi_{c2} \pi^- K_S^0$ and
%    (e)  $B^- \to \chi_{c1} \pi^0 K^+$ decay modes.
%    }
%  \label{fig:MKpi_all_3body}
%\end{figure}

\begin{figure*}%[ht]
  %\centering
  \includegraphics[trim=0cm 1cm 0cm 0.25cm,height=90mm,width=160mm]{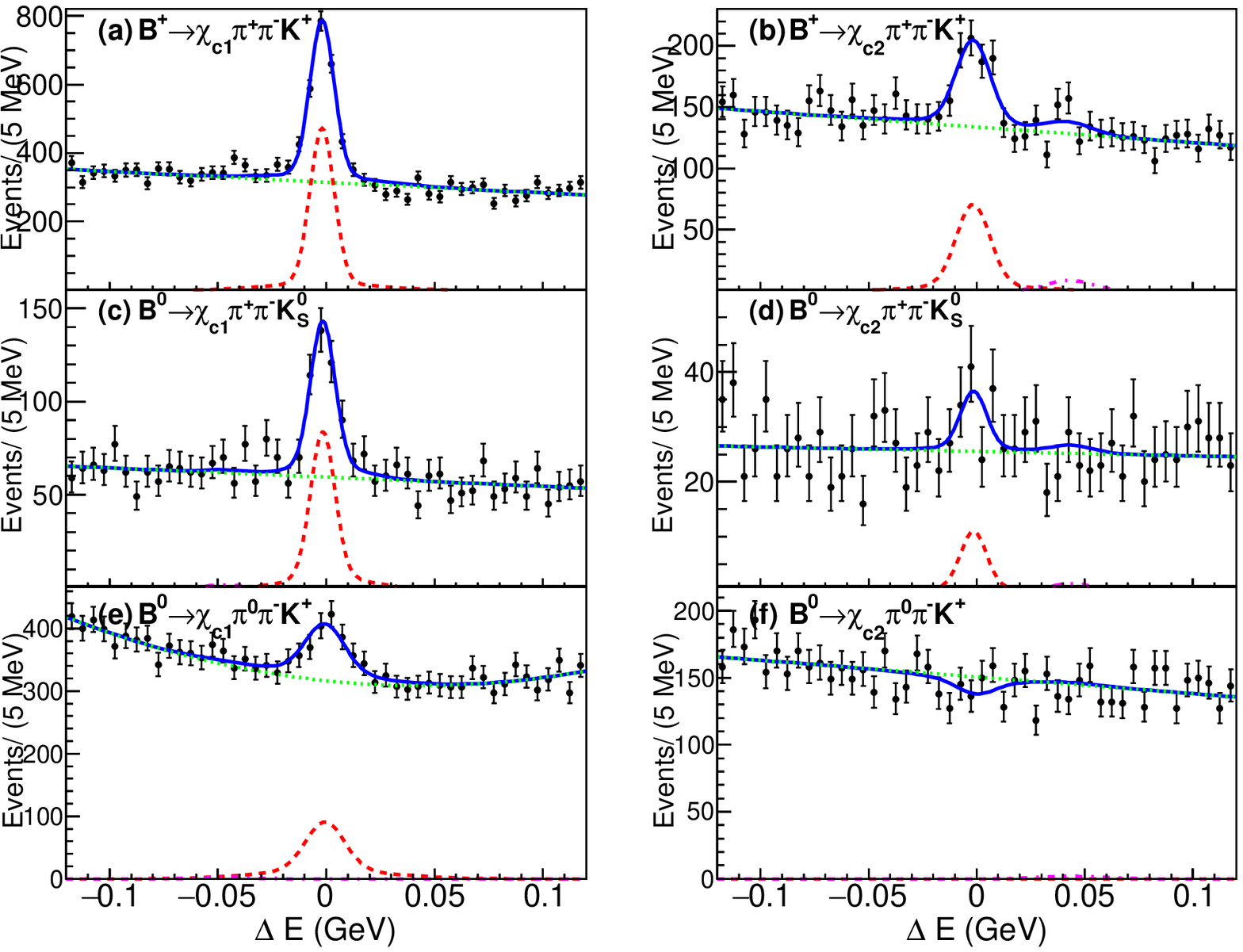}
  \caption{(color online) $\Delta E$ distributions 
    for the (a) $B^+ \to \chi_{c1} \pi^+ \pi^- K^+$,
    (b) $B^+ \to \chi_{c2} \pi^+ \pi^- K^+$,
    (c) $B^0 \to \chi_{c1} \pi^+ \pi^- K_S^0$,
    (d) $B^0 \to \chi_{c2} \pi^+ \pi^- K_S^0$,
    (e) $B^0 \to \chi_{c1} \pi^0 \pi^- K^+$ and
    (f) $B^0 \to \chi_{c2} \pi^0 \pi^- K^+$ decay  modes.
%    (g) $B^- \to \chi_{c1} \pi^0 \pi^- K_S^0$ and
%    (h) $B^- \to \chi_{c2} \pi^0 \pi^- K_S^0$  decay modes. 
    The curves show the signal 
    (red dashed),
    peaking background (magenta dash-dotted) and 
     the background component (green dotted for 
    combinatorial) as well as the overall fit (blue solid).}
  \label{fig:B_chicJ_Kpipi}
\end{figure*}

%%%%
\begin{figure*}%[ht]
  %\centering
  \includegraphics[height=75mm,width=70mm]{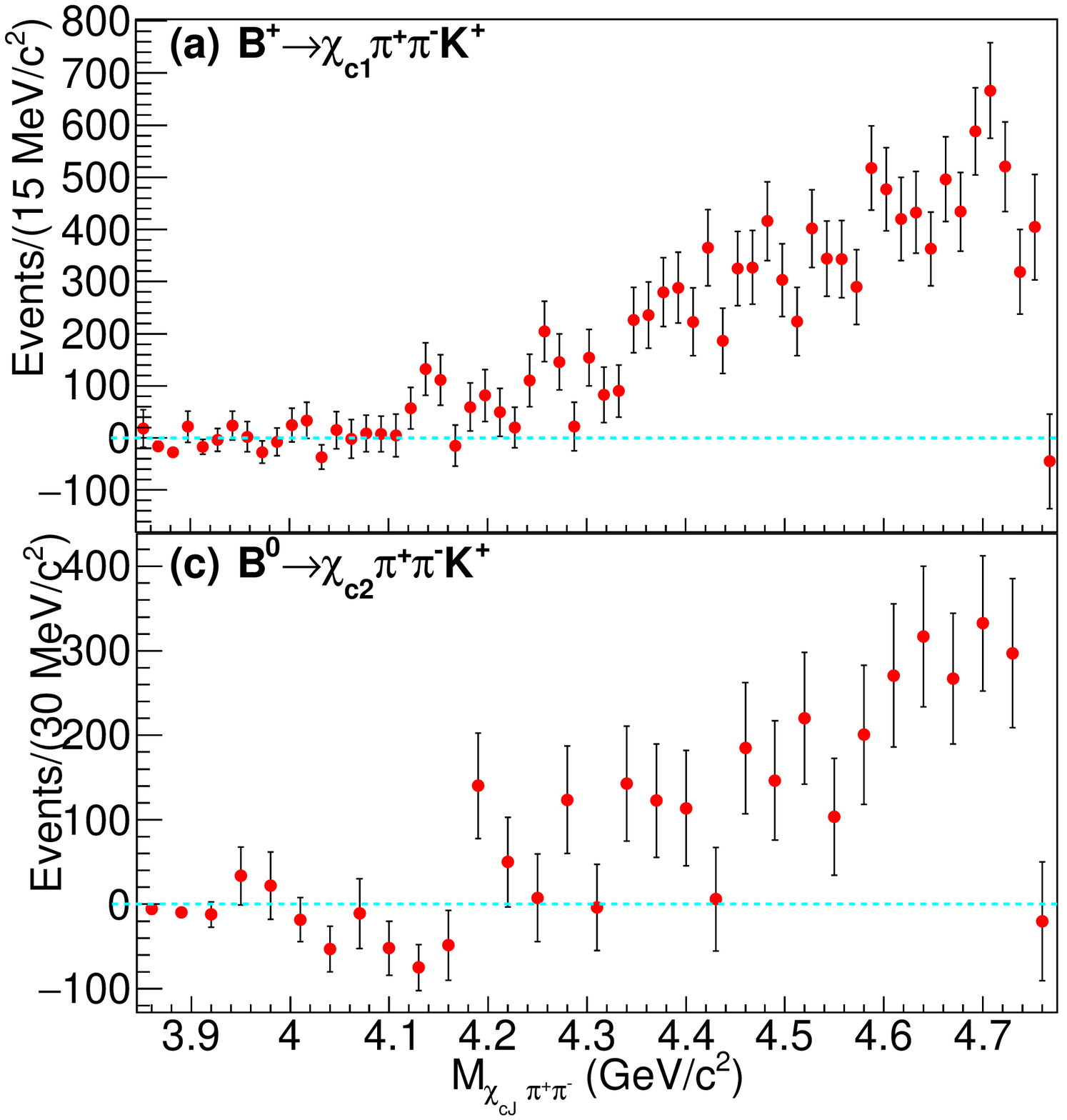}
  \includegraphics[height=75mm,width=70mm]{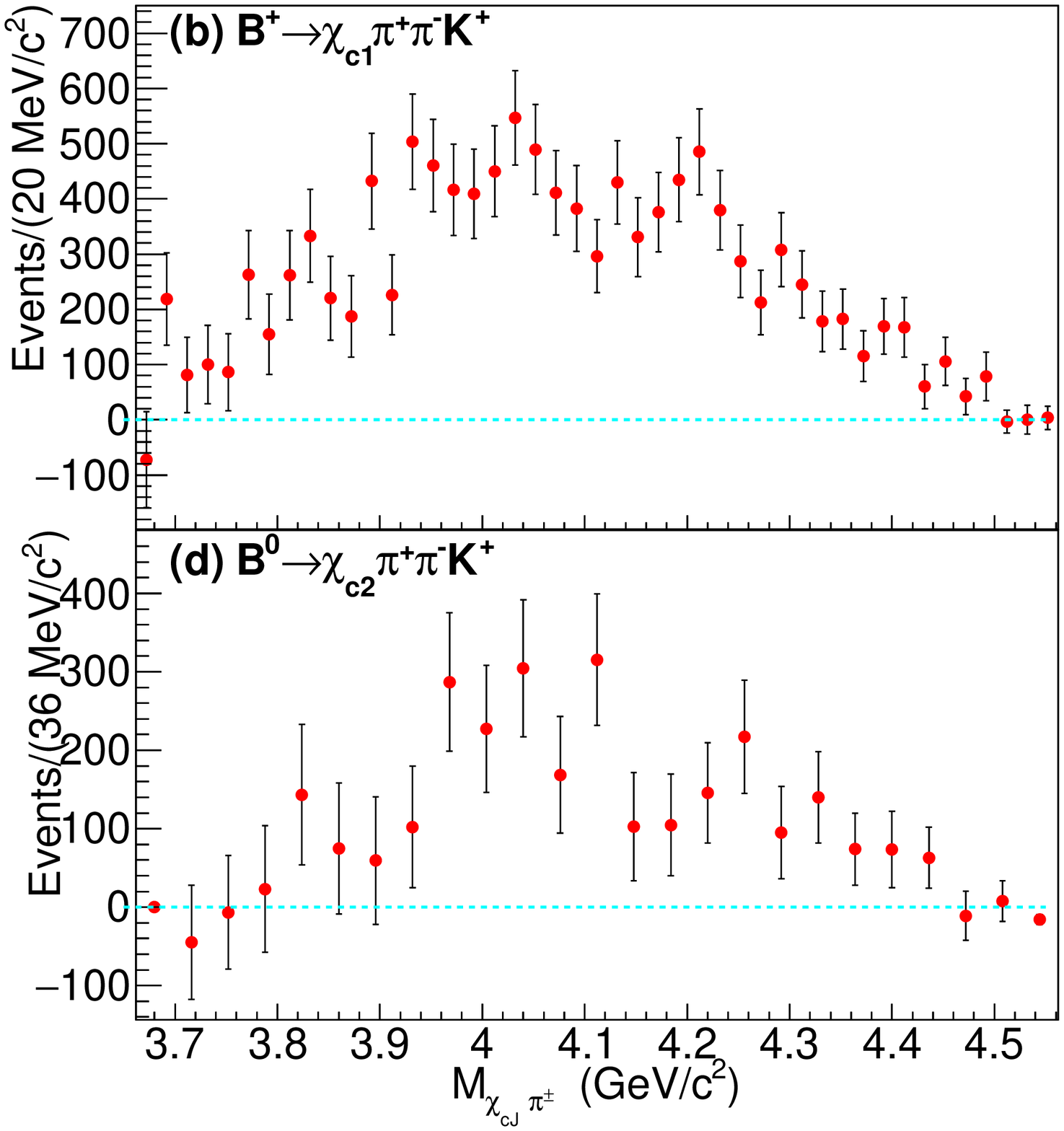}
  \caption{(color online) Background subtracted  efficiency corrected 
  $_{S}\mathcal{P}$\textit{lot}
    (a) $M_{\chi_{c1} \pi^+ \pi^-}$,
    (b) $M_{\chi_{c1} \pi^\pm}$,
    (c) $M_{\chi_{c2} \pi^+ \pi^-}$ and   
    (d) $M_{\chi_{c2} \pi^\pm}$ distributions for the
    $B^+ \to \chicx \pi^+ \pi^- K^+$ decay modes.   
    }
  \label{fig:Mchipi_B_chicJ_Kpipi}
\end{figure*}

%%%%
\begin{figure*}%[ht]

  %\centering
 \includegraphics[height=75mm,width=53mm]{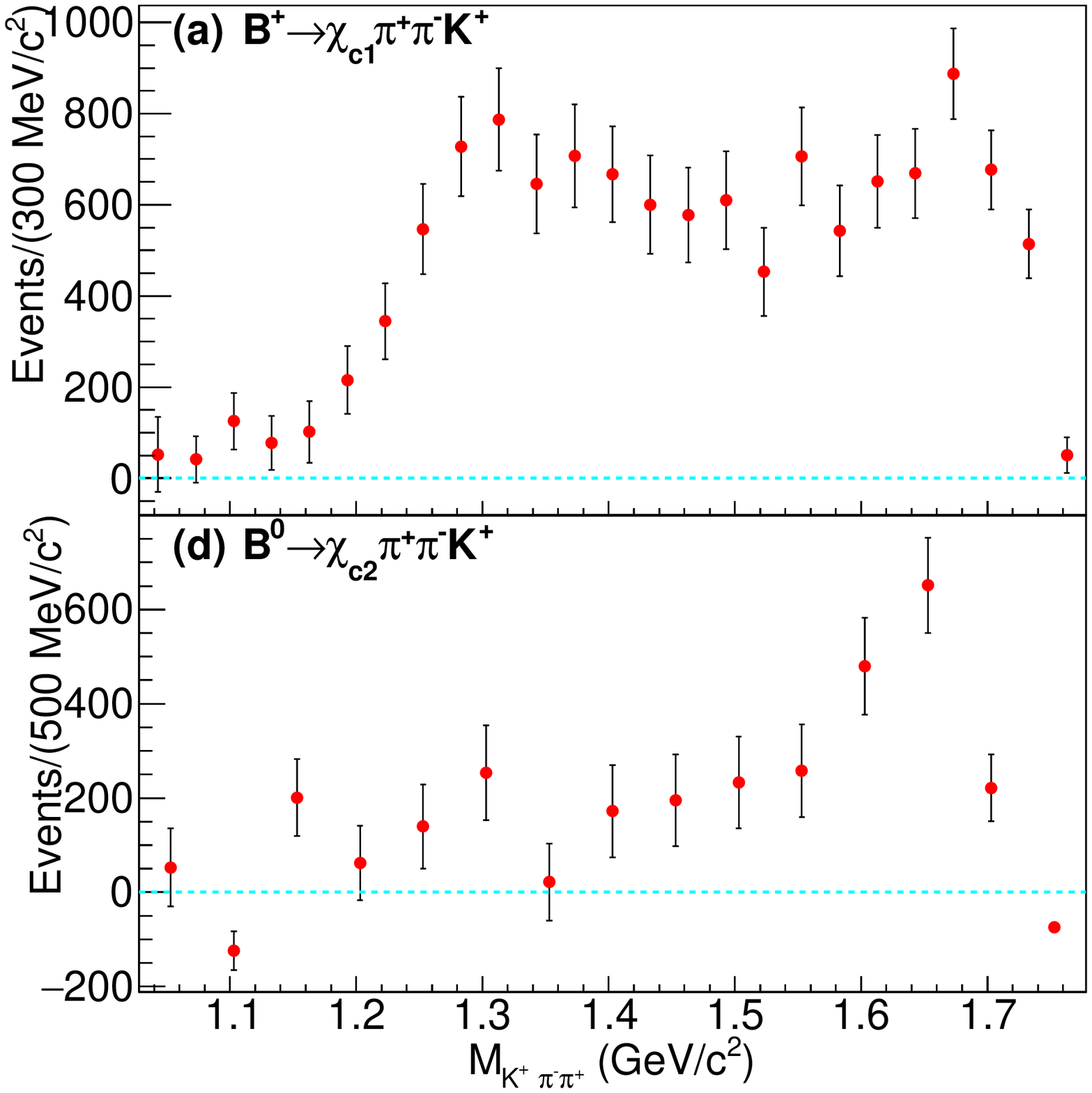}%M_Kpipi_4.eps}
 \includegraphics[height=75mm,width=53mm]{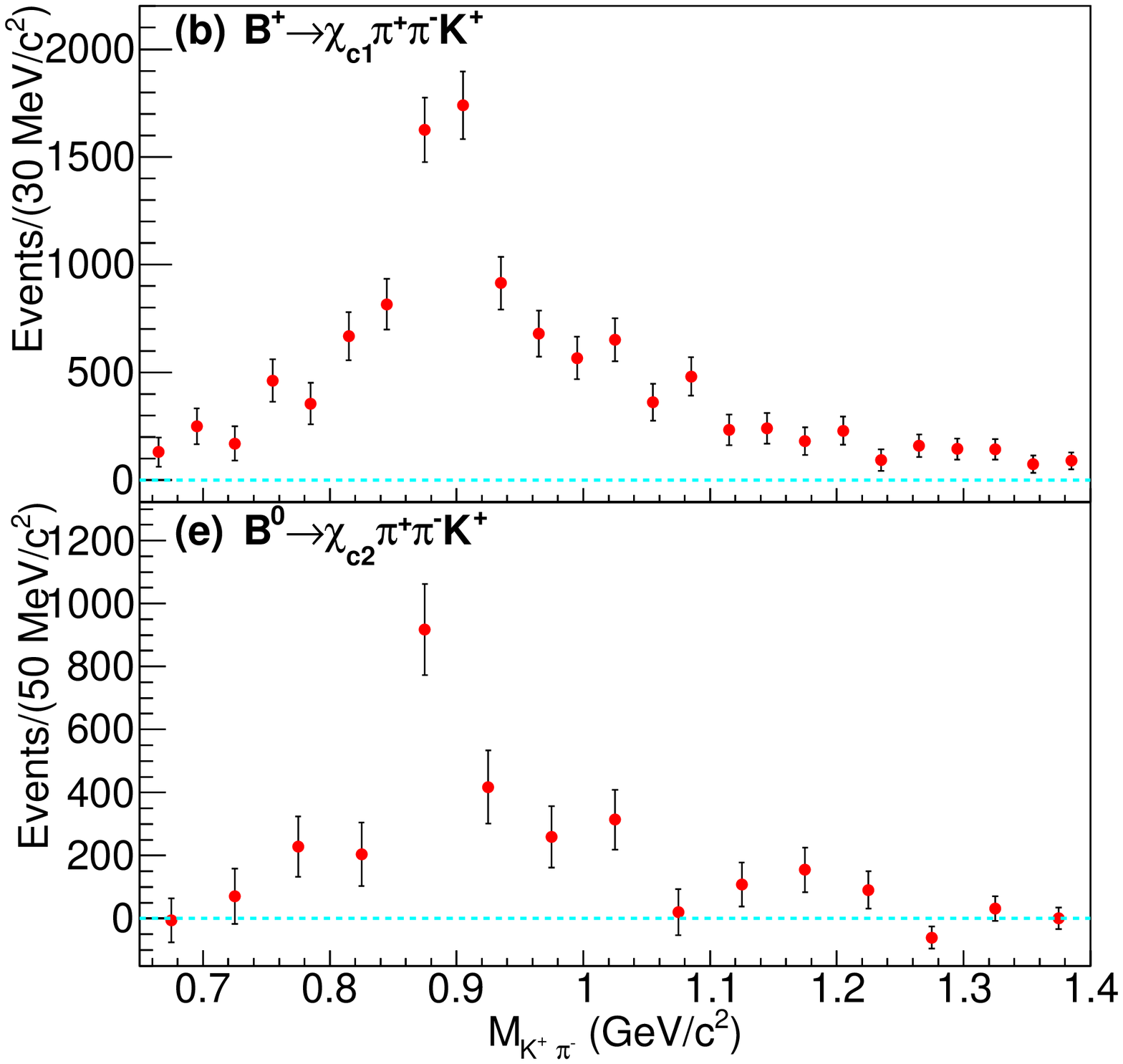}%M_Kpi_4.eps}
 \includegraphics[height=75mm,width=53mm]{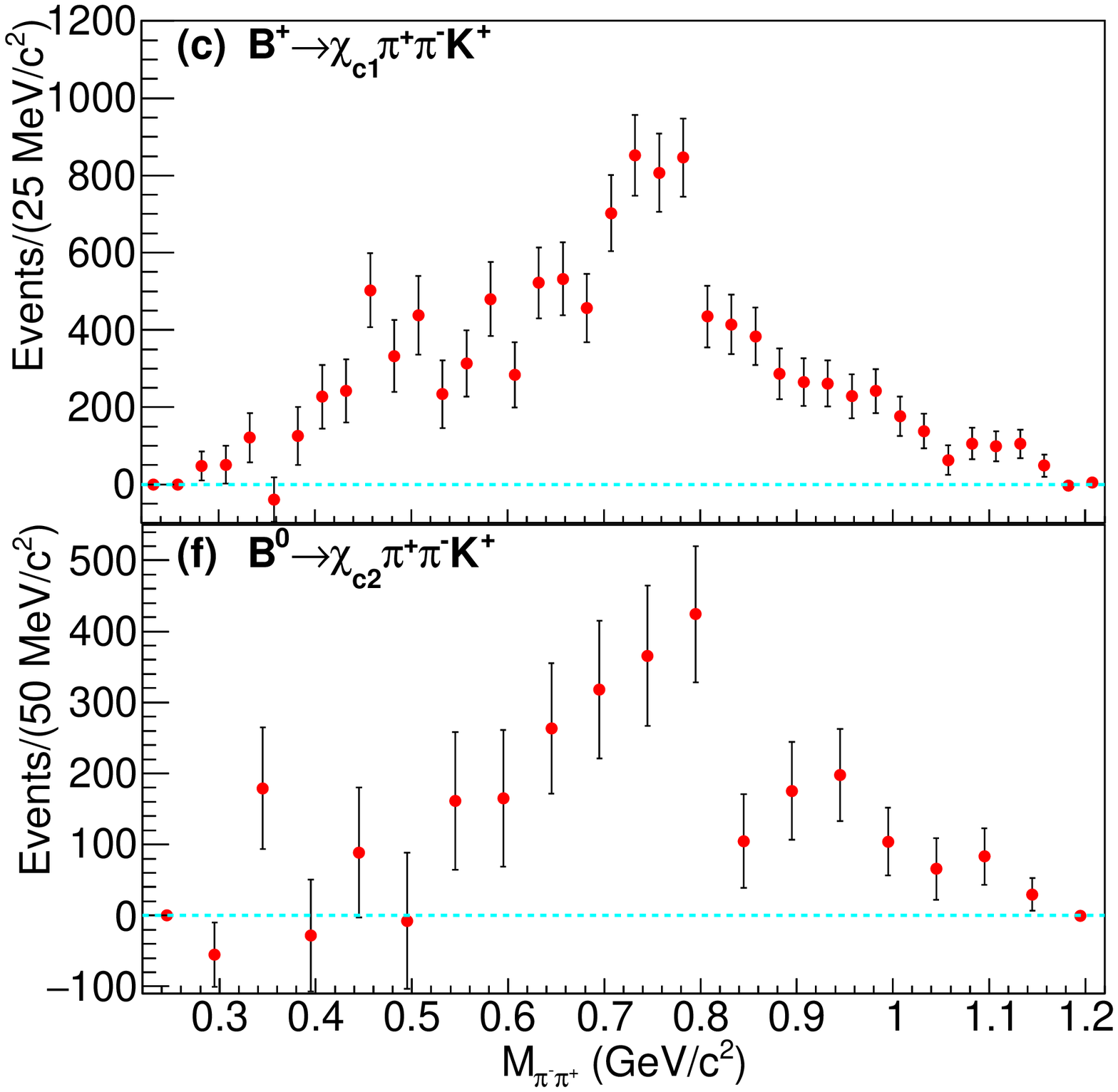}%M_pipi_4.eps}

 \caption{(color online) Background subtracted  
   efficiency corrected
   $_{S}\mathcal{P}$\textit{lot}
    (a and d) $M_{K^+ \pi^+ \pi^-}$,
    (b and e) $M_{K^+ \pi^-}$ and
    (c and f) $M_{\pi^+ \pi^-}$ distributions for 
    $B^+ \to \chi_{c1} \pi^+ \pi^- K^+$ decay (upper) and  
    $B^+ \to \chi_{c2} \pi^+ \pi^- K^+$ decay (lower), respectively.   
    }
  \label{fig:MKpi_B_chicJ_Kpipi}
\end{figure*}

%\begin{figure}[h!]
%  \centering
%  \includegraphics[height=108mm,width=90mm]{./fig/chipipimass_B-chic1_pipi_Ks.eps}
%  \caption{ Background subtracted 
%    (a) $M_{\chi_{c1} \pi^+ \pi^-}$,
%    (b) $M_{\chi_{c1} \pi^+}^{max}$
%    and (c) $M_{\chi_{c1} \pi^+}^{min}$ distribution for
%    $B^0 \to \chi_{c1} \pi^- \pi^+ K_s^0$ decay mode.
%       }
%  \label{fig:Mchipipi_B_chic1pipiKs}
%\end{figure}

%\begin{figure}[h!]
%  \centering
%  \includegraphics[height=108mm,width=90mm]{./fig/Kspipimass_B_chic1_pipi_Ks.eps}
%  \caption{ Background subtracted $M_{K\pi}$ distribution for
%    (a) $M_{K_S^0 \pi^+ \pi^-}$,
%    (b) $M_{K_S^0 \pi^+}^{max}$
%    and (c) $M_{K_S^0 \pi^+}^{min}$ distribution for
%    $B^0 \to \chi_{c1} \pi^- \pi^+ K_s^0$ decay mode.
%    }
%  \label{fig:MKspipi_B_chic1pipiKs}
%\end{figure}

%......................................................................
%......................................................................
% ..  B -> chicx pi- K+
\subsection{ \boldmath $B \to \chicx \pi K$}
To study 
$\chicx$ production in  three-body $B$ decays, we use
charged and neutral kaons and pions to reconstruct the $B$ decay mode
of interest: 
$B^0 \to \chicx \pi^- K^+$,   $B^+ \to \chicx \pi^+ K_S^0$ and  
$B^+ \to \chicx \pi^0 K^+$.
%To combine charged kaon and pion candidate tracks, care for
%opposite charge is taken into account. 
The signal is identified using kinematic requirements on
$\Delta E$ and $M_{\rm bc}$.  Among
the events containing $B$ candidates, 10\%, 16\% and 22\% have
multiple candidates in the $B^0 \to \chicx \pi^- K^+$, 
$B^+ \to \chicx \pi^+ K^0_S$
and $B^+ \to \chicx \pi^0 K^+$ modes,
respectively. The aforementioned BCS procedure is used to select the 
$B$ candidate in such events.

The UML fit to the $\Delta E$ distribution for the $B^0 \to \chicx  \pi^- K^+$
and $B^+ \to \chicx \pi^+ K_S^0$ decay modes is shown in 
Fig.~\ref{fig:B_chicJ_Kpi} (a)-(d).
For $B^+ \to \chicx  \pi^0 K^+$ 
decays, the signal is modeled by the sum of a Gaussian and 
a logarithmic Gaussian~\cite{lg}. For $B^+ \to \chi_{c1} \pi^0 K^+$ 
decays, the mean and width of the core 
Gaussian are floated and the remaining parameters are fixed according to MC;
for $B^+ \to \chi_{c2} \pi^0 K^+$ 
decays, all parameters are fixed after applying the data/MC correction 
estimated from the  $B^+ \to \chi_{c1} \pi^0 K^+$ 
decay mode. No peaking background is expected in the 
$B^+ \to \chi_{c1} \pi^0 K^+$ decay mode while,
 in $B^+ \to \chi_{c2} \pi^0 K^+$, feed-down from 
$B^+ \to \chi_{c1} \pi^0 K^+$  is expected and is modeled by  a Gaussian PDF 
(whose yield  and all parameters  are fixed from MC simulation study). 
The rest of the background is combinatorial and  modeled using a first-order
Chebyshev polynomial. The fit to the $\Delta E$ distribution  for 
$B^+ \to \chicx  \pi^0 K^+$ is shown in Fig.~\ref{fig:B_chicJ_Kpi} (e) and (f). 

We obtain $2774\pm 66$ ($206\pm 25$), $770\pm35$ ($76\pm15$) and
$803\pm 70$ ($17.5\pm28.4$) %and $138.92\pm 23.25$ ($5.68\pm 13.01$) 
signal events for  the $B^0 \to \chi_{c1} \pi^- K^+$ ($B^0 \to \chi_{c2} \pi^- K^+$),
$B^+ \to \chi_{c1} \pi^+ K_S^0$ ($B^+ \to \chi_{c2} \pi^+ K_S^0$) and
$B^+ \to \chi_{c1} \pi^0 K^+$ ($B^+ \to \chi_{c2} \pi^0 K^+$) decay modes %and
%$B^0 \to \chi_{c1} \pi^0 K_S^0$ ($B^0 \to \chi_{c2} \pi^0 K_S^0$) decay mode
having a significance of 67$\sigma$ (8.7$\sigma$),
34$\sigma$ (4.6$\sigma$) and  
16$\sigma$ (0.4$\sigma$), respectively.
% and 6.0$\sigma$ (0.4$\sigma$), respectively.
The significance is estimated using  the  value of
$-2\ln(\mathcal{L}_{0}/\mathcal{L}_{\rm max})$, where 
$\mathcal{L_{\rm max}}$ ($\mathcal{L_{\rm 0}}$) denotes the likelihood value 
when the yield is allowed to vary (is set to zero). 
The systematic uncertainty, which is described below, 
is included in  the significance calculation~\cite{cousinhighland}.
We make the first observation of  the $B^0 \to \chi_{c2} \pi^- K^+ $ decay mode
along with the first evidence for a $B^+ \to \chi_{c2} \pi^+ K_S^0$ decay.
We estimate the branching fractions according to the formula
$\mathcal{B} = Y /(\epsilon \times  \mathcal{B}_s \times N_{B\bar{B}})$; 
here $Y$ is the yield, $\epsilon$ is the reconstruction efficiency, 
$\mathcal{B}_s$ is the secondary branching fraction taken from 
Ref.~\cite{pdg}, and $N_{B\bar{B}}$ is the number of $B \bar{B}$ mesons 
in the data sample. Equal production of neutral and 
charged $B$ meson pairs in the $\Upsilon(4S)$ decay is assumed.
Table~\ref{tab:final_results} summarizes the results.

%Figure~\ref{fig:Mchipi_all_3body} shows the background subtracted 
% $M_{K^+ \pi^-}$ and $M_{\chicx \pi^+}$ distributions. 
The $K^{*}(892)$ is found to be a major contribution in the 
$B \to \chi_{c1} \pi K$ decay modes as seen from 
Fig.~\ref{fig:Mchipi_all_3body} (a), (e) and (i);
in $B \to \chi_{c2} \pi K$ decays, the $K^{*}(892)$ component is less prominent
and a cluster of events around $M_{K^\pm \pi^\mp}$ = 1.4 GeV$/c^2$ 
shows a relatively large contribution.  
Our study suggests that the $B \to \chi_{c2} K^*(892)$ mechanism 
does not 
dominate the $B \to \chi_{c2} \pi K$ decay,   in marked contrast to 
the $\chi_{c1}$ case. Until now,  the previous measurements
of $\chi_{c2}$ ~\cite{BaBar_PRL_102_132001_2009,LHCb_NPB_874_3_663_2013} 
were limited to $B^0 \to \chi_{c2} K^{*}(892)^0$ only and so 
were not able to observe three-body $B$ decays. 
From this study, one may posit that the production mechanism of the 
$\chi_{c2}$  from $B$ mesons is different in 
three-body decays for the  $B \to  \chi_{c1} \pi K $ case.
As shown in Fig.~\ref{fig:Mchipi_all_3body} (b) and (f),
the  $M_{\chi_{c1} \pi^{\pm}}$ distributions
are similar to those obtained by a previous Belle study~\cite{RomanMizuk}
in which a Dalitz analysis suggested two charged $Z$ states decaying into 
$\chi_{c1} \pi^+$. Also, the $M_{\chi_{c1} \pi^0}$ distribution in
Fig.~\ref{fig:Mchipi_all_3body} (j) shows a similar behavior as seen in
the charged $M_{\chi_{c1} \pi^{\pm}}$ distribution.  However,  
due to limited statistics,  no noticeable 
feature in the  $M_{\chi_{c2} \pi^+}$ spectrum is seen as shown 
in the corresponding Fig.~\ref{fig:Mchipi_all_3body} (d) and (h).

In decay modes where we find no significant signal,
we determine a 90\% C.L. upper limit (U.L.) on its branching fraction
with a frequentist method that uses ensembles of pseudo-experiments.
For a given signal yield, 10000 sets of signal and background events are
generated according to their PDFs and fits are performed. The U.L. is determined
from the fraction of samples that give a yield larger than that of data.

%structure is similar to the one obtained by Belle in observation
%of two charged $Z$ states in $B^0 \to \chi_{c1} \pi^- K^+$~\cite{RomanMizuk}.
%However, in $M_{\chi_{c2} \pi^+}$ nothing can be said due to limited statistics. From $M_{K^+ \pi^-}$ distributions, in 
%Figure~\ref{fig:MKpi_all_3body}(a,b), $K^*(892)$ component dominates 
%in $B \to \chi_{c1} K^+ \pi^-$ decay mode while  $K^*(1430)$ (or higher $K^*$
%resonance) is found to be dominant in  $B \to \chi_{c2}  \pi^- K^+$ decay mode.Till now only $B^0 \to \chi_{c2} K^*(892)^0$, was known 
%experimentally~\cite{BaBar_PRL_102_132001_2009,LHCb_NPB_874_3_663_2013}. 
%$B \to \chi_{c1} \pi^-  K^+$ has  a substantial $K^+ \pi^-$ [non-resonant (NR)]
%component while $B \to \chi_{c1}  \pi^-  K^+$ is mostly
%coming from $K^*$ resonance. \textcolor{red}{(can we provide 
%$\chi_{c1} K^*(892)$ alone ? so we can compare to other results...)}

%We didn't performed fit to extract different components of charged $Z$ and 
%$K^*$ resonance as it is outside the scope of this paper. Although, more 
%detailed study  (Dalitz analysis) can be performed in future.

%%First talk  about Delta E. SO, put together

%..................................................c.....................
%B+ -> chicx pi+ pi- K+
\subsection{\boldmath $B \to \chicx \pi \pi K$}
Each $\chicx$ candidate is combined with a pair of 
oppositely charged pions (or a charged-neutral pair)
and a kaon (either $K^\pm$ or $K^0_S$) to reconstruct the
$B$ decays of interest:
$B^+ \to \chicx \pi^+ \pi^- K^+$,  $B^0 \to \chicx \pi^+ \pi^- K_S^0$ and 
$B^0 \to \chicx \pi^- \pi^0 K^+$ decay modes.
% and   $B^+ \to \chicx \pi^+ \pi^0 K_S^0$ decay modes. O
Of the selected $B$ candidates, identified by the 
$\Delta E$ and $M_{\rm bc}$ requirement, 35\%,
35\% and 50\% have multiple candidates
in the $B^+ \to \chicx \pi^+ \pi^-  K^+$,
$B^0 \to \chicx \pi^+ \pi^-  K_S^0$
and
$B^0 \to \chicx \pi^- \pi^0 K^+$ decay modes, respectively.
%and $B^0 \to \chicx \pi^- \pi^0 K^+$ decay mode, respectively.
In case of multiple $B$ candidates, the aforementioned BCS
is used to select a single $B$ candidate in the event.

The signal yield is extracted from a 1D UML fit to the $\Delta E$  
distribution as shown in Fig.~\ref{fig:B_chicJ_Kpipi}. We get
$1502\pm70$ ($269\pm34$),
$268  \pm 30$ ($37.8 \pm 14.2$) and
$545 \pm 81$ ($-76.7 \pm 42.0$) signal events
%$125 \pm 44$ ($82.0 \pm 28.4$) signal events
with  a $19.2\sigma$ ($8.4\sigma$), $7.1\sigma$ ($1.8\sigma$) and
$6.5 \sigma$ (null)  significance %and $2.8 \sigma$ ($2.7 \sigma$) significance
for the
$B^+ \to \chi_{c1} \pi^+ \pi^- K^+$ ($B^+ \to \chi_{c2} \pi^+ \pi^-  K^+$),
$B^0 \to \chi_{c1} \pi^+ \pi^-  K_S^0$ ($B^0 \to \chi_{c2}  \pi^+ \pi^-  K_S^0$)
and
$B^0 \to \chi_{c1} \pi^- \pi^0  K^+$ ($B^0 \to \chi_{c2}  \pi^- \pi^0  K^+$)
%and $B^0 \to \chic_{c1} \pi^- \pi^0 K^+$ ($B^0 \to \chic_{c1} \pi^- \pi^0 K^+$)
decay modes, respectively.
For the first time, we  observe the 
$B^+ \to \chi_{c1} \pi^+ \pi^- K^+$,
$B^+ \to \chi_{c2} \pi^+ \pi^-  K^+$,
$B^0 \to \chi_{c1} \pi^+ \pi^-  K_S^0$,
%$B^0 \to \chi_{c2}  \pi^+ \pi^-  K_S^0$,
and $B^0 \to \chi_{c1} \pi^- \pi^0  K^+$ decay modes.
%We also do find an indication for a signal for the
%$B^0 \to \chicx \pi^- \pi^0 K^+$ decay mode.
Table~\ref{tab:final_results} summarizes the fit results.

In order to understand the dynamics of the production of $\chicx$ in four-body $B$ decays, we examine
the background-subtracted  $_{S}\mathcal{P}$\textit{lot}
distribution of
$M_{\chicx \pi \pi}$,   $M_{\chicx \pi^\pm}$,
$M_{K \pi \pi}$, $M_{K^+ \pi^-}$, 
and $M_{\pi^+ \pi^-}$,  which are shown in
Figs.~\ref{fig:Mchipi_B_chicJ_Kpipi} 
and \ref{fig:MKpi_B_chicJ_Kpipi}
for the $B^+ \to \chicx \pi^+ \pi^- K^+$  decay mode.
No narrow resonance can be seen in the $M_{\chicx \pi^+ \pi^-}$   and
$M_{\chicx \pi^\pm}$ distributions with
the current statistics. 
There seems to be an enhancement of signal events around 4.1-4.2 GeV$/c^2$ 
in $M_{\chicx \pi \pi}$  that is due to cross-feed;
the same effect is seen in our $B \to J/\psi X$ MC sample
that is used to study the background. %From these plot, it is not much clear.
Higher $K^*$ resonances are seen in the $M_{K^{+} \pi^- \pi^{+}}$ and
$M_{K^+ \pi^-}$ distributions shown in 
Fig.~\ref{fig:MKpi_B_chicJ_Kpipi} similar to the ones seen in the
$B^+ \to J/\psi \pi^+ \pi^-  K^+$ decay mode~\cite{Guler}. There is a peaking 
structure near 1680 MeV$/c^2$  due to the
$K^*(1680)^+$. Further, a $K^*(892)^0$ peak is found in
$M_{K^+\pi^-}$.   
Here again, the contrast between $B^+\to  \chi_{c2} \pi^+ \pi^- K^+$ 
decays and those to $\chi_{c1}$ is apparent: the decays to $\chi_{c2}$ 
mostly include higher
$K^*$ resonances.
Figures~\ref{fig:MKpi_B_chicJ_Kpipi} (e) and (f) show the 
$M_{\pi^+ \pi^-}$ distributions
for the $B^+ \to \chicx \pi^+ \pi^-  K^+$ decay mode, 
which suggest a contribution
from $\rho$ as an intermediate state.
%In the light of what we see in $B \to \chicx K \pi$ and 
%$B \to \chicx K \pi \pi$, 

\begin{table*}[!htbp]
\caption{ Summary of the results. Signal yield ($Y$) from the fit, 
significance ($\mathcal{S}$) with systematics included,
corrected efficiency ($\epsilon$) and
measured $\mathcal{B}$.
For $\mathcal{B}$, the first (second) error is statistical (systematic). 
 Here, in the neutral $B$ decay case,  the
$K_S^0 \to \pi^+ \pi^-$ branching fraction  is included
in the efficiency ($\epsilon$) but the factor of 2 
(for $K^0 \to K_S^0~{\rm or}~K_L^0$)
is taken into account separately. ${\cal R}_{\cal B}$ is the ratio of $\mathcal{B}(B \to \chi_{c2} X)$ to $\mathcal{B}(B\to \chi_{c1} X)$, where $X$ is the same 
set of particles accompanying the $\chi_{c1}$ ($\chi_{c2}$) in the
final states.
}
\begin{center}

  \begin{tabular}{lcccccc}
\hline \hline
Decay & Yield ($Y$) & $\mathcal{S} (\sigma)$ & $\epsilon$(\%) & ${\cal B}$ $(10^{-4})$ & ${\cal R}_{\cal B}$  \\ \hline

\multicolumn{5}{l}{$B^0 \to\chicx \pi^- K^+$} & $0.14\pm 0.02$
 \\ \hline
 $\chi_{c1}$ & $2774 \pm 66 $ & 66.7 & 17.9  & $4.97\pm0.12\pm0.28$  & \\ %\multirow{2}{}{$0.15 \pm 0.02$}  \\
 $\chi_{c2}$ & $206 \pm 25 $& 8.7 & 16.2   & $0.72\pm0.09\pm0.05$ & \\ \hline
 
 \multicolumn{5}{l}{$B^+ \to \chicx \pi^+ K^0$}  & $0.20\pm 0.04$  
 \\ \hline
 $\chi_{c1}$ & $770 \pm 35$  &  33.7 &  8.6 &  $5.75\pm 0.26 \pm 0.32$ & \\ % \multirow{2}{}{$0.20 \pm 0.04$} \\
 $\chi_{c2}$ &  $76.4 \pm 14.7 $ &  4.6 & 7.5 & $1.16\pm 0.22\pm0.12$&  \\ \hline

\multicolumn{5}{l}{$B^+ \to\chicx \pi^0 K^+$} &   $<0.21$
\\ \hline
$\chi_{c1}$ & $803 \pm 70$ & 15.6 & 7.8 & $3.29\pm0.29\pm0.19$ & \\% \multirow{2}{}{$<0.21$} \\
$\chi_{c2}$ & $17.5\pm 28.4 $ & 0.4 & 7.0 & $<0.62$ & \\ \hline

%\multicolumn{6}{l}{$B^0 \to \chicx \pi^0 K^0$}  
%\\ \hline
%$\chi_{c1}$ & $139 \pm 23$  &  6.0 &  3.67 &  $2.38 \pm 0.40 \pm 0.20$ & \multirow{2}{}{$<0.61$} \\
%$\chi_{c2}$ &  $5.7 \pm 13.0 $ &  0.4 & 3.35 & $<1.19$ & \\ \hline

\multicolumn{5}{l}{$B^+ \to \chicx \pi^+ \pi^- K^+$}  & $0.36\pm0.05$   %$\mathcal{B}(10^{-6})$  
\\ \hline
$\chi_{c1}$ & $1502 \pm 70$ & 19.2 & 12.8 & $3.74 \pm 0.18 \pm0.24 $ & \\% \multirow{2}{}{$0.36\pm 0.05$} \\
$\chi_{c2}$ & $269 \pm 34$ & 8.4 & 11.4 & $1.34\pm0.17\pm0.09$ & \\ 
\hline

\multicolumn{5}{l}{$B^0 \to \chicx \pi^+ \pi^- K^0$}  & $<0.61$
\\ \hline
$\chi_{c1}$ & $268 \pm 30 $ &  7.1 & 5.4 &  $3.16\pm0.35\pm0.32$ & \\% \multirow{2}{}{$<0.59$} \\ 
$\chi_{c2}$ &  $37.8 \pm 14.2 $ & 1.8 & 4.8 &  $<1.70$  & \\ 
\hline

\multicolumn{5}{l}{$B^0 \to \chicx \pi^- \pi^0 K^+$}  & $<0.25$ 
\\ \hline
$\chi_{c1}$ & $545 \pm  81$ &6.5 & 5.0  & $3.52 \pm 0.52 \pm0.24$ & \\% \multirow{2}{}{$<0.25$} \\
$\chi_{c2}$ & $-76.7 \pm 42.0 $ & --- & 4.3 & $<0.74$  &\\  \hline
\hline
%\multicolumn{6}{l}{$B^+ \to \chicx \pi^0 \pi^+ K^0$}  
%\\ \hline
%$\chi_{c1}$ & $125 \pm 44$ &  2.8 & 2.14 &  $<5.38$  & \multirow{2}{}{--} \\ 
%$\chi_{c2}$ &  $82.0 \pm 28.4 $ &  2.7 & 1.87 &  $<7.12$ &  \\ 
%\hline \hline

\end{tabular}
\label{tab:final_results}
\end{center}
\end{table*}

\subsection*{Search for {\boldmath $X(3872)$} and {\boldmath $\chi_{c1}(2P)$}}
\begin{figure}[h!]
  \centering
  \includegraphics[trim=-0.5cm 0cm 0cm -1.0cm,height=50mm,width=80mm]{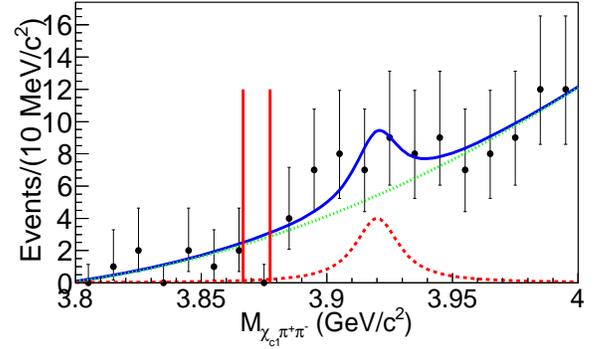}
  \caption{ (color online) The $\chi_{c1} \pi^+ \pi^-$ invariant mass 
    spectrum for $B^+ \to \chi_{c1} \pi^+ \pi^- K^+$ candidates. Two vertical 
    red lines show the $\pm 3 \sigma$ window to search for $X(3872) \to \chi_{c1} \pi^+ \pi^-$. The curves show the
$\chi_{c1}(2P)$ signal (red dashed) and the background
    (green dotted) and the overall fit (blue solid). }  
  \label{fig:X3872search}
\end{figure}

To search for the  $X(3872) \to \chi_{c1} \pi^+ \pi^-$, we investigate
the signal in the $M_{\chi_{c1} \pi^+ \pi^-}$ distribution within the 
signal-enhanced window
of $-20$ MeV $< \Delta E <$ 20 MeV for 
$B^+ \to \chi_{c1} \pi^+ \pi^- K^+$ candidates. In the
absence of any 
significant peak as shown in Fig.~\ref{fig:X3872search},
we count the number of events within the $\pm 3 \sigma$ window and
find no events. Therefore, we use 2.6 events as the upper limit
of the signal yield based on the Feldman and Cousins approach~\cite{FC_UL}
including systematic uncertainty of the detection efficiency.
Using $5.6\%$  as the corrected efficiency
for $B^+ \to X(3872)(\to \chi_{c1} \pi^+ \pi^-) K^+$ estimated
from signal MC, we obtain
$\mathcal{B}(B^{+}\to X(3872)K^{+}) \times 
\mathcal{B}(X(3872)\to\chi_{c1}\pi^+\pi^-) < 1.5 \times 10^{-6}$ ($90\%$ C.L.).

The $\chi_{c1} (2P)$ signal in the $M_{\chi_{c1}\pi\pi}$ spectrum is described by a
PDF composed as the convolution of a Breit-Wigner function with a Gaussian. 
As a plausible assumption for the $\chi_{c1}(2P)$ state,  its
mass and width are fixed at 3920 MeV/$c^2$ and 20 MeV, assuming the
PDG interpretation of $X(3915)$=$\chi_{c0}(2P)$ and
the property of  $\chi_{c2}(2P)$~\cite{pdg}.
The width of the Gaussian is fixed to $2~{\rm MeV}$, corresponding
to the detector resolution in the mass estimation obtained from MC
simulated samples.
The fit (shown in Fig.~\ref{fig:X3872search}) results
in a signal yield of $12.2 \pm 9.1$ events, which 
translates to 
$30.3$ events  at the $90\%$ confidence level. A product branching 
fraction upper limit is extracted, including statistical and 
systematic uncertainties and the $8.9\%$ reconstruction efficiency: $\mathcal{B}(B^+\to\chi_{c1}(2P)K^+)\times \mathcal{B}(\chi_{c1}(2P)\to\chi_{c1}(1P)\pi^+\pi^-) < 1.1 \times 10^{-5}$ ($90\%$ C.L.).

Table~\ref{tab:UL} summarizes our search for $X(3872)$ and $\chi_{c1}(2P)$ 
in the $B^+ \to (\chi_{c1} \pi^+ \pi^-) K^+$ decay mode.

\begin{table}[htbp!]
\caption{\label{tab:UL} U.L. for
  $B^+ \to X(\to \chi_{c1} \pi^+ \pi^-) K^+$; here $X$ stands for
  $X(3872)$ and the assumed $\chi_{c1}(2P)$. The upper limit at 
  (90\% C.L.) includes
  the systematics ($N^{U.L.}$), corrected efficiency ($\epsilon$)  and
  product of branching fractions
  $\mathcal{B}(B^+\to\ X K^+)\times \mathcal{B}(X\to\chi_{c1}\pi^+\pi^-)$ 
  ($\mathcal{B}^{U.L.}$).}

\begin{tabular}{lccc}
  \hline \hline  
  Mode & $~~~Y^{U.L.}~$ & $~~\epsilon$ (\%)~ &  $~~~\mathcal{B}^{U.L.}$ ($\times 10^{-5}$) \\ \hline
  $X(3872)$  & $< 2.6$  &  5.6 &$ < 0.15$  \\
  $\chi_{c1}(2P)$ & $<30.3$  & 8.9 & $<1.10$ \\
\hline \hline
\end{tabular}
\end{table}

%.......................................................................

%.
%%%%%%%%%%%
%%

\subsection{Systematics}
Table~\ref{tab:syserrall} summarizes the systematic for each mode.
Corrections for small differences in the signal detection efficiency 
between MC and data have been applied for the lepton  and kaon 
identification requirements,  as was done in the inclusive study. 
In addition to the items commonly affecting the inclusive branching
fractions measurements, we consider the following systematic
uncertainty sources.
In  Belle, dedicated
$D^{*+} \to D^0(K^- \pi^+)\pi^+$  samples are used 
to estimate the kaon (pion) identification efficiency correction.  
To estimate  the correction  and 
residual systematic uncertainty  for $K_S^0$  reconstruction, 
$D^{*+}\to D^0(\to K_S^0 \pi^+ \pi^-) \pi^{+}$ samples are used. 
For $\pi^0$, the efficiency correction and systematic uncertainty are estimated 
from a sample of $\tau^- \to \pi^+ \pi^0 \nu_\tau$ decays. The errors on 
the PDF shapes are obtained by varying all fixed parameters by $\pm 1 \sigma$
and taking the change in the yield as the systematic uncertainty. 
%The uncertainties due to the secondary branching fractions are also taken into
%account.  The 
%uncertainty on the photon identification  and the difference of $\pi^0$ veto
%between data and MC is applied to be same as that in inclusive $\chicx$ 
%measurement study.

\begin{table*}[!htbp]
  \caption{\label{tab:syserrall} Summary of  
      systematic uncertainties on  the $B \to \chicx X$ branching fraction.  
      Uncertainty on lepton identification ($\ell$), kaon identification ($K$), 
      pion identification ($\pi$),  tracking, gamma identification 
      ($\gamma$ id), $K_S^0$ reconstruction,  $\pi^0$ reconstruction,
      $\pi^0$ veto, uncertainty in the secondary branching fractions, 
      PDFs used to extract signal yield
      and  uncertainty on the $N_{B\bar{B}}$.  
  }

 \begin{tabular}{lcccccccccccc|c}
    \hline \hline
    & \multicolumn{12}{c}{Uncertainty (\%) } \\
    Mode & $\ell$ &  $K$  & $\pi$ & Tracking & $\gamma$ id & 
    Secondary $\mathcal{B}$ & $K_S^0$ & $\pi^0$  & 
    $\pi^0$ veto & $\epsilon$ &  PDF &   $N_{B\bar{B}}$ &Total \\ \hline

    \multicolumn{13}{l}{$B^0 \to\chicx \pi^- K^+$} %$\mathcal{B}(10^{-4})$ 
    \\ \hline
    $\chi_{c1}$ & 2.1 & 1.0 & 1.0 & 1.4 & 2.0 & 3.6 & ---&---&1.2& 1.0 & 1.4 & 1.4& 5.6 \\ 
    $\chi_{c2}$ &  2.1 & 1.0 & 1.0 &  1.4 & 2.0 & 3.7 & --- &---&1.2& 2.4 & 3.2 & 1.4&6.7 \\ \hline
    
    \multicolumn{13}{l}{$B^+ \to \chicx \pi^+ K^0$}   
    \\ \hline
    $\chi_{c1}$  & 2.1 & ---& 1.0 & 1.8& 2.0 & 3.6 & 0.7 & --- & 1.2 & 1.0 & 0.7& 1.4&5.5  \\
    $\chi_{c2}$  & 2.1 & ---& 1.1 & 1.8& 2.0 & 3.7 & 0.7 & --- & 1.2& 2.2 & 9.1 & 1.4&10.8 
    \\ \hline

    \multicolumn{13}{l}{$B^+ \to\chicx \pi^0 K^+$} %$\mathcal{B}(10^{-4})$  
    \\ \hline
    $\chi_{c1}$ & 2.1& 1.0& ---& 1.1& 2.0& 3.6& ---& 2.2& 1.2& 1.0& 1.7&1.4& 5.9 
    \\
    $\chi_{c2}$ & 2.1& 1.0& ---& 1.1& 2.0& 3.7& ---& 2.2& 1.2& 2.4& 191& 1.4&191.1 
    \\ \hline

    \multicolumn{13}{l}{$B^+ \to \chicx \pi^+ \pi^- K^+$}    %$\mathcal{B}(10^{-6})$  
    \\ \hline
    $\chi_{c1}$& 2.1& 1.1& 2.3& 1.8& 2.0& 3.6& ---& ---& 1.2& 1.0& 2.6&1.4& 6.5
    \\
    $\chi_{c2}$& 2.1& 1.2& 2.3& 1.8& 2.0& 3.7& ---& ---& 1.2& 2.2& 2.0&1.4& 6.6
    \\
    \hline
    
    \multicolumn{13}{l}{$B^0 \to \chicx \pi^+ \pi^- K^0$} 
    \\ \hline
    $\chi_{c1}$& 2.1& ---& 2.3& 2.1& 2.0& 3.6& 0.7& ---& 1.2& 1.0& 8.1&1.4& 10.1
    \\
    $\chi_{c2}$& 2.1& ---& 2.3& 2.1& 2.0& 3.7& 0.7& ---& 1.2& 2.3& 31.5&1.4& 32.1  
    \\
    \hline

    \multicolumn{13}{l}{$B^0 \to \chicx \pi^- \pi^0 K^+$} %$\mathcal{B}(10^{-6})$  
    \\ \hline
    $\chi_{c1}$& 2.1& 1.1& 1.1& 1.4& 2.0& 3.6& ---& 2.2&  1.2& 1.1& 3.6&1.4& 6.9
    \\
    $\chi_{c2}$& 2.1& 1.2& 1.1& 1.4& 2.0& 3.7& ---& 2.2& 1.2& 2.9& **&1.4& 7.5
    \\ 
    
    \hline \hline

    \hline \hline
  \end{tabular}

\end{table*}

\subsection{Discussion on Exclusive decays}
Table~\ref{tab:final_results} summarizes the studied exclusive decays of 
$B$ to  $\chicx X$ decays. For the first time, we observe the
$B^0 \to \chi_{c2} \pi^- K^+$,
$B^+ \to \chi_{c2} \pi^+ K_S^0$, $B^+ \to \chi_{c2} \pi^+ \pi^- K^+$,
$B^+ \to \chi_{c1} \pi^+ \pi^- K^+$,
$B^0 \to \chi_{c1} \pi^+ \pi^- K^0$ and
$B^0 \to \chi_{c1} \pi^- \pi^0 K^+$ decay modes. 
%We also saw a hint of $B^+ \to \chicx \pi^+ \pi^0 K^0$ decay modes.
We find that  in  three-body decays the $\chi_{c2}$
is more likely to be produced in association with higher 
$K^*$ resonances; in contrast, decays to $\chi_{c1}$ are
accompanied predominantly by 
the $K^*(892)$.
The same  phenomenon is  observed in the four-body 
production of $\chi_{c2}$ and 
$\chi_{c1}$ from $B$ decays.  No strong hint for any narrow resonance 
(less than 5 MeV width) is seen 
in the $M_{\chicx \pi}$ and $M_{\chicx \pi \pi}$ distributions. 
If one adds the measured branching fraction in this paper
(excluding the obtained U.L.), we obtain 
$\mathcal{B}(B \to \chi_{c1}  n \pi K)$ with $n \in \{1,2\}$
to be $(1.75 \pm 0.08)\times 10^{-3}$, which corresponds to
a $(58 \pm 5)\%$ fraction of the measured $\mathcal{B}(B \to \chi_{c1} X)$.
Using $\mathcal{B}(B^+ \to \chi_{c1} K^+)$~\cite{pdg},  this
accounts for $(74 \pm 6)\%$ of  $B$ mesons decaying into $\chi_{c1} X$.
Similarly, $\mathcal{B}(B \to \chi_{c2}  n \pi K)$ with 
$n\in\{1,2\}$ is $(0.23 \pm 0.02)\times 10^{-3}$,
corresponding to  $(32\pm 5)\%$ of the inclusive 
$\mathcal{B}(B \to \chi_{c2} X)$.
For the treatment of the uncertainty, no correlation is assumed and
the uncertainty is the sum of the systematic and statistical uncertainties 
in quadrature. 

%One can clearly see that the sum of exclusively measured decay modes
%of $B \to \chi_{c1} X$ occupies (73$\pm$6)\% of the inclusive 
%$\mathcal{B}(B \to \chi_{c1}X)$ for direct $\chi_{c1}$ production in 
%$B$ decays, 
%while   $B \to \chi_{c2} X$ measured decay modes cover 
%one-third of the decay modes.

%.......................................................................
\section{Summary}

We measured the feed-down-contaminated
$\mathcal{B}(B\to \chi_{c1} X)$ and $\mathcal{B}(B\to \chi_{c2}X)$ of
$(3.33\pm 0.05\pm 0.24)\times 10^{-3}$ and 
$(0.98\pm 0.06\pm0.10)\times 10^{-3}$, respectively, 
where the first (second) error is statistical (systematic).
After subtracting the $\psi'$ feed-down contributions, we find the 
pure inclusive branching fractions 
$\mathcal{B}(B\to \chi_{c1} X)$ and
$\mathcal{B}(B\to \chi_{c2}X)$ of
$(3.03 \pm 0.05\pm 0.24)\times 10^{-3}$ and
$(0.70\pm 0.06\pm0.10)\times 10^{-3}$, respectively.
Here, the systematic uncertainty dominates. 
For inclusive production of $\chicx$, we measure the ratio
$\mathcal{B}(B \to \chi_{c2} X) /\mathcal{B}(B \to \chi_{c1} X)$ of 
$(23.1 \pm 2.0 \pm 2.1)\%$. 
We observe the $B^0 \to \chi_{c2} \pi^- K^+$ decay mode 
for the first  time, with $206\pm25$ signal events and 
a significance of 8.7$\sigma$,  along with evidence for the 
$B^+ \to \chi_{c2} \pi^+ K_S^0$ decay  mode, with  76$\pm$15 signal events
and a significance of $4.6\sigma$.
In four-body decays, we observe the 
$B^+ \to \chi_{c1} \pi^+ \pi^- K^+$, $B^+ \to \chi_{c1} \pi^+ \pi^- K^+$,
$B^0 \to \chi_{c1}  \pi^+ \pi^- K_S^0$,  and
$B^0 \to \chi_{c1}  \pi^0 \pi^- K^+$ decay modes for the first time 
and report on measurements 
of their branching fractions.  We find that  
$\chi_{c2}$ production, in contrast with $\chi_{c1}$,
increases with a higher number of multi-body $B$ decays:
$\mathcal{R}_{\mathcal{B}}$ 
for $B^+ \to \chicx \pi^+ \pi^- K^+$ decay (0.36$\pm$0.05) is almost twice
that measured in the $B^0 \to \chicx \pi^- K^+$ decay mode ($0.20\pm0.04$).
We observe that the $\chi_{c2}$ is more often accompanied by 
higher $K^*$ resonances, in contrast to the $\chi_{c1}$ 
that is dominantly produced with the lower 
$K^*$ resonance. All previous 
studies~\cite{BaBar_PRL_102_132001_2009,LHCb_NPB_874_3_663_2013}
were limited 
to $K^*(892)^0$, while our study suggests that $\chi_{c2}$ 
is preferentially produced
with higher $K^*$ resonances. Clearly, to study $\chi_{c2}$ 
production in $B$ decays,  it is important to avoid considering
solely the lower $K^*$ 
resonances. Suppression in two-body $B$ decays is found 
to be due
to the factorization hypothesis~\cite{bdecays_fact}.  
In our search for $X(3872) \to \chi_{c1} \pi^+ \pi^-$ and
$\chi_{c1}(2P)$, we determine an U.L. on 
the product of branching fractions 
$\mathcal{B}(B^+ \to X(3872) K^+) \times (X(3872) \to \chi_{c1} \pi^+ \pi^-)$
[$\mathcal{B}(B^+ \to\chi_{c1}(2P) K^+) \times (\chi_{c1}(2P) \to \chi_{c1} \pi^+ \pi^-)$]   $<$  $1.5 \times 10^{-6}$ [$1.1 \times 10^{-5}$] at  the
90\% C.L. The 
negative result for our searches is compatible with the interpretation of
$X(3872)$ as an admixture state of a 
$D^0 \bar{D}^{*0}$ molecule and a $\chi_{c1}(2P)$ charmonium state.

\acknowledgments

\section{Acknowledgments}
We thank the KEKB group for the excellent operation of the
accelerator; the KEK cryogenics group for the efficient
operation of the solenoid; and the KEK computer group,
the National Institute of Informatics, and the 
PNNL/EMSL computing group for valuable computing
and SINET4 network support.  We acknowledge support from
the Ministry of Education, Culture, Sports, Science, and
Technology (MEXT) of Japan, the Japan Society for the 
Promotion of Science (JSPS), and the Tau-Lepton Physics 
Research Center of Nagoya University; 
the Australian Research Council;
Austrian Science Fund under Grant No.~P 22742-N16 and P 26794-N20;
the National Natural Science Foundation of China under Contracts 
No.~10575109, No.~10775142, No.~10875115, No.~11175187, and  No.~11475187;
the Chinese Academy of Science Center for Excellence in Particle Physics; 
the Ministry of Education, Youth and Sports of the Czech
Republic under Contract No.~LG14034;
the Carl Zeiss Foundation, the Deutsche Forschungsgemeinschaft
and the VolkswagenStiftung;
the Department of Science and Technology of India; 
the Istituto Nazionale di Fisica Nucleare of Italy; 
the WCU program of the Ministry of Education, National Research Foundation (NRF) 
of Korea Grants No.~2011-0029457,  No.~2012-0008143,  
No.~2012R1A1A2008330, 
No.~2013R1A1A3007772, 
No.~2014R1A2A2A01005286,
No.~2014R1A2A2A01002734, 
No.~2015R1A2A2A010032\\80, No.~2015H1A2A1033649;
the Basic Research Lab program under NRF Grant No.~KRF-2011-0020333,
Center for Korean J-PARC Users, No.~NRF-2013K1A3A7A06056592; 
the Brain Korea 21-Plus program and Radiation Science Research Institute;
the Polish Ministry of Science and Higher Education and 
the National Science Center;
the Ministry of Education and Science of the Russian Federation and
the Russian Foundation for Basic Research;
the Slovenian Research Agency;
the Basque Foundation for Science (IKERBASQUE) and 
the Euskal Herriko Unibertsitatea (UPV/EHU) under program UFI 11/55 (Spain);
the Swiss National Science Foundation; the National Science Council
and the Ministry of Education of Taiwan; and the U.S.\
Department of Energy and the National Science Foundation.
This work is supported by a Grant-in-Aid from MEXT for 
Science Research in a Priority Area (``New Development of  
Flavor Physics''),  for Scientific Research on Innovative
Areas (``Elucidation of New Hadrons with a Variety of Flavors''),
and from JSPS for Creative Scientific 
Research (``Evolution of Tau-lepton Physics'').


\begin{thebibliography}{99}
\bibitem{CX}
Hereinafter, $\chicx$ refers to  either $\chi_{c1}$ or $\chi_{c2}$, 
depending  on  which is reconstructed.
\bibitem{dCB}
The double-sided Crystal Ball function is defined as :\\
$f(x) = N_0 
\left(n_l/|\alpha_l| \right)^{n_l}  \exp \left( - |\alpha_l|^2/2 \right) \times
\\~~~~~~~~~~~~~~~~~~~~~~~~~~~~~\left [n_l/|\alpha_l| - |\alpha_l| -  (x - \mu)/\sigma\right]^{-n_l},
\\~~~~~~~~~~~~~ \text{if}~ (x-\mu)/\sigma \leq -\alpha_l  
\\~~~~~~~~=~ N_0 \left( n_r/|\alpha_r| \right)^{n_r}  \exp \left( - |\alpha_r|^2/2 \right) \times
\\~~~~~~~~~~~~~~~~~~~~~~~~~~~~~\left [n_r/|\alpha_r| - |\alpha_r| + (x - \mu)/\sigma\right]^{-n_r}, 
\\~~~~~~~~~~~~~ \text{if}~ (x-\mu)/\sigma \geq \alpha_r 
\\~~~~~~~=~ N_0 \exp \left[ - (x - \mu)^2/(2\sigma^2) \right],  ~~\text{otherwise}.
$\\
Here $N_0$ is the normalization, $\sigma$ is the standard deviation, $\mu$ is the mean,
$n_l$ ($n_r$), $\alpha_l$ ($\alpha_r$) are shape parameters for left (right)
tail.
\bibitem{mixchg}
Charge-conjugate and neutral modes are included throughout the paper 
unless stated otherwise.
\bibitem{lg}
The logarithmic Gaussian is parameterized as:\\
$f(x)=N_0/c \times$ $\exp{- \{\ln[(\epsilon-x)/(\epsilon-x_p)]\}^2/(2\sigma^2_0)}$  
%$\exp{- {\ln[(\epsilon - x)/(\epsilon - x_p)]}^2}/(2\sigma_0^2)}$,
where $\epsilon = \sigma/a + x_p$, $c=\sqrt{2\pi}\sigma_0(\epsilon -x)$ and 
$\sigma_0 = (\ln(a\sqrt{2\ln 2} + \sqrt{1 + 2a^2\ln 2})/\sqrt{2\ln 2}$.  
Here, $N_0$ is the  normalization, $\sigma$ is the standard deviation, 
$x_p$ is the mean and $a$ is the asymmetry.
\end{thebibliography}

\end{document}